\documentclass[preprint,12pt,numbering]{elsarticle}

\usepackage{graphicx}
\usepackage{ulem}
\usepackage{amssymb}
\usepackage{xcolor}
\usepackage{soul}
\sethlcolor{yellow} 
\usepackage{tcolorbox}
\usepackage{float}
\usepackage{arydshln}
\usepackage{pifont}      
\usepackage{booktabs}
\usepackage{tabularx}
\usepackage{arydshln}    
\usepackage{amssymb}   
\usepackage{pifont}      
\usepackage{float}      
\usepackage{array}       
\usepackage{booktabs}   
\usepackage{amssymb}    
\usepackage{array}       

\tcbuselibrary{breakable}
\usepackage{amsmath}

\usepackage{subcaption}

\journal{Journal of Computational Physics}

\begin{document}

\begin{frontmatter}

\title{Fast and accurate  quasi-atom method
for simultaneous atomistic and continuum simulation of solids}

\author[label1]{Artem Chuprov\corref{cor1}}
\author[label1]{Egor Nuzhin}
\author[label2]{Alexey Tsukanov}
\author[label1,label3]{Nikolay Brilliantov}

\cortext[cor1]{Corresponding author}
\address[label1]{Skolkovo Institute of Science and Technology, Bolshoy Boulevard 30, bld. 1, Moscow, 121205, Russia}
\address[label2]{Schmidt Institute of Physics of the Earth of the Russian Academy of Sciences B. Gruzinskaya str., 10, bld. 1, Moscow, 123242, Russia}
\address[label3]{School of Computing and Mathematical Sciences, University of Leicester, Leicester LE1 7RH, UK}
\ead{nikolaigran@gmail.com}

\begin{abstract}
We report a novel hybrid method of simultaneous atomistic simulation of solids in  critical regions (contacts surfaces, cracks areas, etc.), along with continuum modeling of other parts. The continuum is treated in terms of quasi-atoms of different size, comprising composite medium. The parameters of interaction potential between the quasi-atoms are  optimized to match elastic properties of the composite medium to those of the atomic one. 
The optimization method coincides conceptually  with the online Machine Learning (ML) methods, making it  computationally very efficient. Such an approach allows a straightforward application of standard software packages for molecular dynamics (MD), supplemented by the ML-based optimizer. The new method is  applied to model systems with a simple, pairwise Lennard-Jones potential, as well with multi-body Tersoff potential, describing covalent bonds. Using LAMMPS software we simulate collision of particles of different size. Comparing simulation results, obtained by the novel method, with full-atomic simulations, we demonstrate its accuracy, validity and overwhelming superiority in computational speed. Furthermore, we compare our method with other hybrid methods, specifically, with the closest one -- AtC (Atomic to Continuum) method. We demonstrate a significant superiority  of our approach in computational speed and implementation convenience. Finally, we discuss a possible extension of the method for modeling other phenomena.

\end{abstract}









\end{frontmatter}

\section{Introduction}
\label{sec1}

Molecular dynamics (MD) simulations have become an indispensable tool for studying materials at the atomic scale, with several software packages, such as LAMMPS (Large‐scale Atomic/Molecular Massively Parallel Simulator) \cite{tutor2024lammps, parks2008peridynamics}, GROMACS \cite{lindahl2001gromacs}, NAMD \cite{phillips2005scalable}, and HOOMD-blue \cite{glaser2015strong}, providing ever-faster engines and GPU‐accelerated implementations to optimize the performance. These frameworks enable precise simulations of complex phenomena, from atomic bond vibrations to collective material behavior, making MD a reliable instrument for probing mechanical properties, phase transitions, and dynamic processes at the molecular level \cite{kumar2024improved}. 

Despite an exponential increase in computing power, fully atomistic simulations of large‐scale systems remain computationally prohibitive \cite{muser2023}. To overcome this constraint, mesoscale and hybrid approaches have been developed \cite{munozbasagoiti2025, kumar2019constitutive, delafrouz2018coarse}, allowing treatment of different regions at different resolution. Nonetheless, many processes such as crack propagation in solids \cite{buehler2008atomistic}, nanoparticle aggregation \cite{hahn2015colloid}, sintering, self-assembly,   dendritic growth \cite{karma1998quantitative}, or phase‐separation dynamics \cite{onuki2002phase} demand simultaneous atomic and continuum description to capture both local chemistry and long‐range morphology. That is, one needs to describe simultaneously large areas as continuum and relatively small regions (critical regions) at the atomic level 
 \cite{espanol2017, nguyen2022, grassl2010}. Traditional continuum or purely coarse-grained methods often fail in this case.

Here, we propose a novel fast and accurate hybrid MD–mesoscale method implemented via a Python–Optimization–LAMMPS bridge. By using full-atomic description for critical regions and representing other regions as built up of “quasi-atoms”, our framework drastically reduces the computational cost, while preserving atomistic accuracy where it is needed. Briefly, quasi-atoms are coarse-grained units (parcels of medium)    encompassing many atoms, defined in  detail below.  
An optimization part of the program automatically calibrates quasi-atom interaction potentials,  to match target elastic moduli, ensuring seamless coupling across scales. In contrast to other hybrid approaches,  the macroscopic characteristics of the continuum  naturally result from the parameters of the  of  model.

To obtain the required quasi-atom potentials in the most effective  way,  we utilize a well-established  optimization method by Regis and Shoemaker \cite{regis2007}, conceptually coinciding with the online Machine Learning, see e.g. \cite{PARISI201954}. We adapt it for the current problem and improve by implementing a parallel sampling strategy, described below. This dramatically increases the computational efficiency of the approach.

Several mesoscopic strategies exist -- Dissipative Particle Dynamics (DPD) \cite{hoogerbrugge1992}, Smoothed Particle Hydrodynamics (SPH) \cite{monaghan1992smoothed}, coarse‐grained force fields \cite{marrink2007martini}, or Adaptive Resolution Schemes (AdResS) \cite{praprotnik2008multiscale}; all these approaches are typically applied to fluids or soft matter, and may not faithfully reproduce the required 
properties of crystalline solids. One can also mention a number of hybrid methods, 

such as e.g. coarse‐grained MD \cite{izvekov2005multiscale},  the heterogeneous multiscale method (HMM) \cite{weinan2003heterogeneous}, or quasi-continuum (QC) methods \cite{tadmor1996mixed,MillerTadmor2002_QC_review, Ortner2008_QC_analysis}.  
 Most of the existing schemes, however,  either target liquids or gases, or suffer from solver overhead, complex setup, and limited applicability to dynamical systems, see the discussion in what follows. 
 Contrarily, our approach addresses crystalline solids -- elasticity, fracture simulations, etc., where many methods can be hardly applied to describe systems' dynamics, while the new method may be straightforwardly implemented.
  
\begin{figure}[t]
  \centering
  \includegraphics[width=\textwidth]{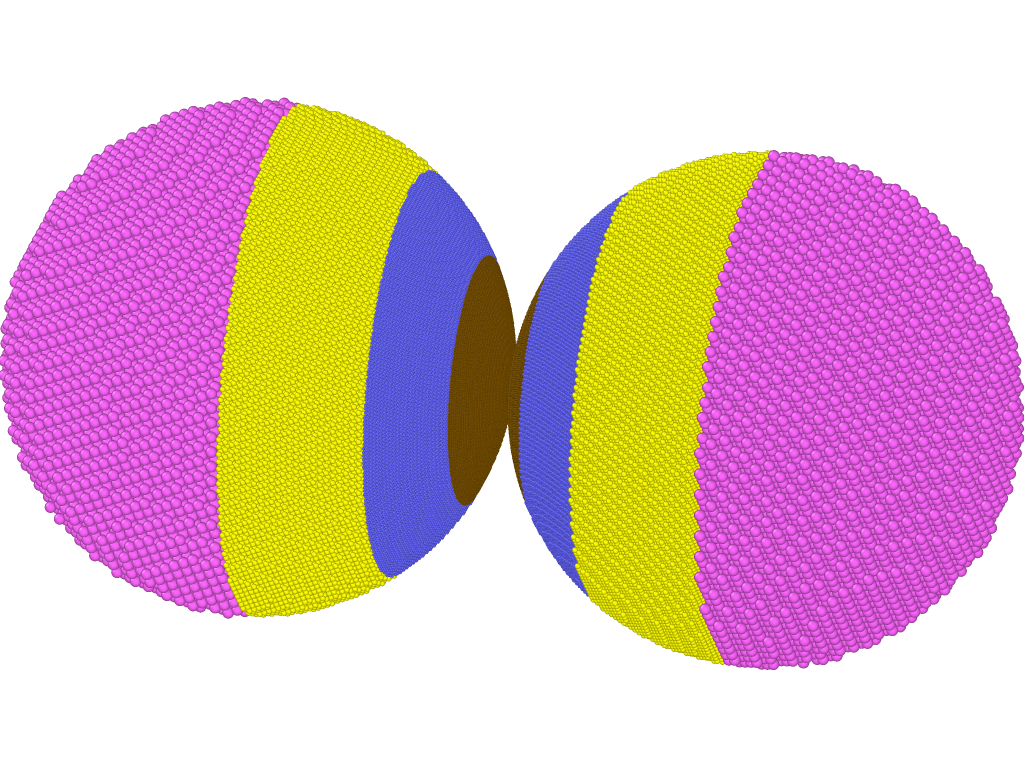}
  \caption{Illustration of two colliding multi-scale particles comprised of real atoms (black) in the critical region - the area of their contact and three different types of quasi-atoms, blue, yellow and rose, which are twice, four and eight times larger than the real atoms.}
  \label{fig:particles_scene}
\end{figure}

For the illustrative purposes we explore here  a  collision of mesoscopically large particles (Fig.~\ref{fig:particles_scene}), however our approach may be readily extended to the problems of crack growth, indentation, and other  multi-scale  problems of solid mechanics.

By integrating LAMMPS --  a high-performance C++ MD engine,  with our flexible Python-based optimizer, our hybrid method leverages LAMMPS’ computational performance for mesoscopic modeling. This overcomes computational limitations of full-atomistic simulations and opens new avenues for multiscale solid mechanics studies. The complete model is available in GitHub via the link\footnote{A. Chuprov, MultiscaleOptimizer 
(https://github.com/ArtemChuprov/MultiscaleOptimizer) (2025)}. 

The rest of the article is organized as follows. In the next Sec. II we describe our hybrid  model, and in Sec. III we present the simulation results for collisions of particles of different size and composition. Here we compare full-atomic and hybrid simulations and verify macroscopic collision theories -- Hertz and JKR theory. Sec. IV is devoted to the comparison of our method with other hybrid method and in the last Sec. V we summarize our findings. All technical details of the new approach are presented in the Appendix.

\section{Description of the hybrid model}
\subsection{Quasi-atom concept}  
\label{subsec1.0} 
 A multi-scale modelling  of a system implies its  simultaneous two-level description -- the continuum one for the  main part and atomistic for smaller part,  where such a description is needed; we call this area "critical region".  
 While the continuum modelling may be performed  by standard numerical techniques, e.g., final elements or final volume, the construction of a smooth transition between the atomistic  and continuum parts is challenging. In the finite volume method, a tessellation of  a system into volume elements is performed and  the laws of continuum mechanics are implemented through the  inter-elements interactions. These interactions are quantified using elastic (viscoelastic) properties of the continuum; they also depend on the size and shape of the volume elements.  A particular tessellation may be done arbitrarily -- it is usually chosen subject to computational convenience.
 
Such a view on the finite volume technique motivates our hybrid approach. Indeed, since any partition with appropriately interacting elements represents the same continuum, one can use  a convenient partition into particles, interacting  with relevant forces, which guarantees the requested properties of the medium. This leads us to the concept of quasi-atoms -- effective particles representing continuum. They may be of different size, which allows a smooth transition from the atomistic part of a system to continuum part: The quasi-atoms' size can gradually grow 
 from the nearly atomic one, in the area, adjacent to the  critical region to mesoscopic/macroscopic size in the bulk of  continuum. 

 Physically, a quasi-atom is not a  collection of real atoms, but a  parcel of medium. It functions as a dynamic discretization node for  continuum region, possessing mass and effective radius, which may be  significantly larger than that of a single atom. In contrast to standard coarse-graining, where beads map to specific molecular clusters, here the quasi-atom is an effective medium unit: its interaction potentials are not derived from local chemical matching, but are engineered to reproduce macroscopic elastic response of a solid. Such an approach provides an  important benefit -- the continuum regions may be  simulated using standard Molecular Dynamics solvers, along with the critical atomistic regions. Furthermore, a smooth transition between these two regions may be easily implemented. 
\subsection{Potentials and Parameters}  
\label{subsec1}  
 In the core of our hybrid multi-scale approach is the usage of the effective MD solvers for both -- atomistic and  continuum parts of the system. For the former part the standard inter-atomic potentials are applicable, for the latter one the appropriate interaction potentials for the quasi-atoms are to be constructed. 
To simulate the behavior of a system, we employ LAMMPS software package for atomic/molecular dynamics simulations. To illustrate the approach, we begin with one of the simplest interatomic potentials -- the Lennard-Jones (LJ) potential:  
\begin{equation}  
\label{LJ}
    U_{LJ}(r) = 4\epsilon \left[ A \left( \frac{\sigma}{r} \right)^{12} - B \left( \frac{\sigma}{r} \right)^{6} \right],  
\end{equation}  
where \( r \) is the interatomic distance, \( \sigma \) defines the zero-potential distance, and \( \epsilon \) sets the energy scale. The LJ potential has  
the repulsive (\(\propto r^{-12}\)) and  attractive  (\(\propto r^{-6}\)) parts. The standard form corresponds to \( A = B = 1 \); setting \( B = 0 \) yields a modified LJ potential with purely repulsive interactions (commonly used in coarse-grained systems).  

Here we validate our method for two representative examples -- for the LJ potential and Tersoff potential, used for materials with covalent bonds \cite{tersoff1988new}. The  approach, however,  is generalizable for other potentials, such as e.g.  Morse potential (for metallic systems) \cite{morse1929diatomic},  machine-learning-driven  potentials \cite{behler2007generalized,bartok2010gaussian},  etc.

Let our system comprises real atoms of the initial size and initial interaction potential and $N-1$ types of the appropriately re-scaled "quasi-atoms". In the case of LJ potential we have two parameters per each atom and quasi-atom type: $(\sigma_i, \epsilon_i)$, with $i=0,\ldots N-1$.  The cross-interaction parameters can be found from the Lorentz-Berthelot rules \cite{delhommelle2001suitability}:

\begin{equation}
\label{LBr}
\begin{aligned}
    \sigma_{ij} &= \frac{\sigma_i + \sigma_j}{2} \\
    \epsilon_{ij} &= \sqrt{\epsilon_i \epsilon_j} \, .
\end{aligned}
\end{equation}

\subsection{Construction of the hybrid crystal lattice}
We simulate a copper material using the Lennard-Jones potential with the  parameters $\epsilon$ = 0.415 eV and $\sigma$ = 0.238 nm, which are close to those of Ref. \cite{venkattraman2012molecular}.  Copper possesses a face-centered cubic (fcc) crystal lattice. Naturally, this is a simplified model of copper, but our primary goal is the methodological validation of the approach, rather than precise material modelling. Our method is not restricted by a specific choice of the initial atomistic potential, which is also confirmed by the simulations with the Tersoff potential, presented below. 

Our approach involves scaling of the material constituents  from the original atoms to much larger (up to e.g. micrometer-size)  quasi-atoms, increasing their size by some factor at each step. Here we use the factor of two, although this can be any reasonable quantity providing smooth transition from the atomistic to continuum part. Hence, in our case, each subsequent type of quasi-atom has an effective diameter  twice larger than that of the previous type and, consequently, twice larger lattice constant and eight times larger mass, 
see Fig.~\ref{fig:particles_scene}.

Given that we know the effective diameter, $\sigma_i$,  for each type of quasi-atoms and have all of the parameters for the initial real atoms, the required number of parameters we need  to find for $(N-1)$ types of quasi-atoms in case of LJ potential will be $(N-1)$. The same idea holds for any other potential -- the equilibrium-distance parameter can be obtained directly from the initial lattice configuration, so no optimization for this quantity is required.

\subsection{Computation of the potential constants for quasi-atoms}

Our goal is to find appropriate constants describing interaction potential between different types of quasi-atoms. Such constants should guarantee that the elastic properties of the effective medium, built up of quasi-atoms, coincide with those of the medium, built of real atoms. This problem is computationally very challenging, since one needs to optimize a series of parameters, by performing expensive MD simulations for each set of constants. Such a class of problems is  called  "high-expensive black-box optimization",  e.g. \cite{CHEN2025117521}. Among the main difficulties there is a lack of knowledge about a gradient in the parametric space, which significantly impedes the  optimization. Fortunately, there exist a powerful tool for solving such problems -- the so-called surrogate optimization. Conceptually it coincides with the online Machine Learning approach \cite{PARISI201954}: a loss function, depending on parameters of interest is introduced and iteratively minimized, with the aim of achieving a global minimum. Referring to Ref. \cite{regis2007} and Appendix for detail,  we briefly sketch the  main ideas here.

To assess the quality of the required parametric  set, we construct a scalar metric -- the loss function, characterizing a  discrepancy between physical properties of the real media (i.e. comprised of real atoms) and the composite one (i.e. comprised of atoms and quasi-atoms). Then we perform an optimization procedure, which minimizes the loss function. In practice, we act as follows. Firstly, we perform full-atomic simulations and obtain the target elastic tensor components $\mathcal{E}_{target}^{ij}$. Secondly, we compute the corresponding tensor $\mathcal{E}_{sim}^{ij}$ for the composite medium (i.e. with quasi-atoms),  for the initial parametric set. Thirdly, we define the loss function as the mean relative error across these tensor components. This ensures that the optimization minimizes the relative deviation rather than the absolute difference, equalizing the weights of stiff and compliant directions. Fourthly,  the loss function is minimized, providing the next iteration for the parametric set. Using the updated set, the second, third and fourth steps are repeated until the set convergence. In what follows we provide key details of the surrogate optimization.

Let $\mathcal{E}_{\text{target}}^{ij}$, with $i,j=x,y,z$ be true  elastic moduli  of the material which we need to obtain by an appropriate choice of the potential constants $\epsilon_k$ and $\sigma_k$ for the $k$-th type of quasi-atoms  ($k=1, \ldots, N-1$). For the cross-interaction potentials between different types of atoms and quasi-atoms we apply Lorentz-Berthelot rules, Eq. \eqref{LBr}. For simplicity we consider interaction only between quasi-atoms of the neighboring sizes, that is, between $(k-1)$-th and $k$-th as well as between $k$-th and $(k+1)$-th types of quasi-atoms. Let $\mathcal{E}_{\text{sim }}^{ij}(\{\epsilon_k,\sigma_k\})$ be the corresponding elastic moduli obtained in MD simulations with the set of potential constants $\{\epsilon_k,\sigma_k\} \equiv (\epsilon_1,\sigma_1, \ldots \epsilon_{N-1},\sigma_{N-1})$, then one needs to find such a set $\{\epsilon_k,\sigma_k\}$, that minimizes the difference:

\begin{equation}
  \label{eq:L_pq1}
  L\bigl(\mathbf\{\epsilon_k,\sigma_k\}\bigr)
  = \frac{1}{6}
  \sum_{p,q \,\in\,\{x,y,z\}}
    \frac{\bigl|\mathcal{E}_{\mathrm{target}}^{pq}
      - \mathcal{E}_{\mathrm{sim}}^{pq}(\mathbf\{\epsilon_k,\sigma_k\}\bigr)\bigr|}
         {\mathcal{E}_{\mathrm{target}}^{pq}} \,.
\end{equation}

The function $L\left( \{\epsilon_k,\sigma_k\}\right)$ is called the "loss function". Ideally, $L\left( \{\epsilon_k,\sigma_k\}\right)$ should be equal to zero  to guarantee that the elastic properties of the effective medium, built up of quasi-atoms, are  exactly equal to these of the real medium. In practice, however, the loss value  differs from zero and one needs to find the best set  $\{\epsilon_k,\sigma_k\}$, that  makes $L\left( \{\epsilon_k,\sigma_k\}\right)$ as small as possible. 

The straightforward computation of the loss function, defined in Eq. \eqref{eq:L_pq1}, is computationally very costly, since full-scale MD simulations for each set $\{\epsilon_k,\sigma_k\}$ are required. Therefore an application of a surrogate loss function $\hat{L}\left( \{\epsilon_k,\sigma_k\}; \theta \right)$ has been proposed \cite{regis2007}. The particular functional form of $\hat{L}$ may be  chosen from the computation convenience; here we use the model of radial basis functions (RBF), see the Appendix for detail. The set of parameters $\theta$ is to be found to provide the best approximation for the actual loss function, $L$, by the surrogate one, $\hat{L}$. In such an approach the search of the optimal set, $\{\epsilon_k^*, \sigma_k^*\}$,  is performed for the analytical surrogate loss function, which requires significantly less computational efforts. Then the quality of the found set $\{\epsilon_k^*,\sigma_k^*\}$ may be checked using the true loss function, with the subsequent improvement of the surrogate loss function. These steps are to be repeated, eventually leading  to the convergence of the true and surrogate loss functions and resulting in the best parametric set $\{\epsilon_k^*,\sigma_k^*\}$.   

Schematically, the following optimization steps are to be performed. Firstly, $n$ parametric sets  $\{\epsilon_k,\sigma_k\}$ are  used to obtain $n$ loss values $L_i \left( \{\epsilon_k,\sigma_k\}\right)$, with $i=1,\ldots n$. Then these $n$ loss values are exploited to find the best surrogate loss function, with the best set of parameters $\theta^*_0$, which may be expressed mathematically  as 
\begin{equation}
    \label{4}
\theta^*_0= {\rm arg \, min}_{\theta} \sum_{i=1}^n \left( L_i\left( \{\epsilon_k,\sigma_k\}\right) - \hat{L}\left( \{\epsilon_k,\sigma_k\}; \theta \right) \right)^2 .  
\end{equation}
This yields the best surrogate loss function $\hat{L}\left( \{\epsilon_k,\sigma_k\}; \theta^*_0 \right)$ at the initialization step. Using the latter function, which is an analytical one, one can  easily find the desired optimal set $\{\epsilon_k^*,\sigma_k^*\}_t$, that is, 
\begin{equation}
    \label{5}
    \{\epsilon_k^*,\sigma_k^*\}_t= {\rm arg \, min}_{\{\epsilon_k,\sigma_k\}}\hat{L}\left( \{\epsilon_k,\sigma_k\}; \theta^*_t \right), 
\end{equation}
 where the subscript $t$ specifies the iteration step ($t=0$ for the initialization step). Next, using $\{\epsilon_k^*,\sigma_k^*\}_t$, we compute the  true  loss value, $L(\{\epsilon_k^*,\sigma_k^*\}_t)$,  again performing the full-scale MD. This allows to improve the surrogate model, by adding a next new point, which may be written as, 
\begin{equation}
    \label{6}
\theta^*_{t+1}= {\rm arg \, min}_{\theta} \sum_{i=1}^{n+t+1} \left( L_i\left( \{\epsilon_k,\sigma_k\}\right) - \hat{L}\left( \{\epsilon_k^*,\sigma_k^*\}_t; \theta \right) \right)^2 ,   
\end{equation}
where the loss values $L_i\left( \{\epsilon_k,\sigma_k\}\right)$ for $i>n$ correspond to $L_i$ with the successive sets $\{\epsilon_k^*,\sigma_k^*\}_0, \, \{\epsilon_k^*,\sigma_k^*\}_1,\ldots \{\epsilon_k^*,\sigma_k^*\}_i,\ldots$. Then the  steps \eqref{5}-\eqref{6} are iteratively repeated, including more and more simulation results into the loss function. This permanently improves the surrogate loss function, until the convergence of the set  $\{\epsilon_k,\sigma_k\}_t$ to its optimal values is achieved. Such an optimization procedure, with the surrogate loss function, allows a significant minimization of costly full-scale MD simulations. 

Note that the above discussion illustrates  only  the main ideas of the procedure, while in practice some additional tricks are employed. For instance, to improve the convergence of the method, a supplementary term is used in the r.h.s. of Eq. \eqref{5}, to enhance an exploration (that is, a search) in the parametric space of $\{\epsilon_k,\sigma_k\}$.  This allows sampling of  areas in the parametric space, where it was too sparse at previous steps. 

Furthermore, we observe that a more accurate optimization may be achieved with additional interaction potentials for the quasi-atoms. Namely, we add harmonic interactions between successive in size quasi-atoms, that is, between their $k$-th and $(k+1)$-th type:
\begin{equation}
    \label{harm}
    U_{harm, k}(r) =\frac12 \varkappa_k \left(r-r_{0,k} \right)^2.
\end{equation} 
Here $r$ is the distance between the quasi-atoms in $k$-th and $(k+1)$-th layer and $r_{0,k}$ is the according equilibrium distance.  For simplicity, we take into account interactions between the closest quasi-atoms of $k$-th and $(k+1)$-th type. Thus, in practical computations we exploited the extended parameteric sets $\{ \epsilon_k,\sigma_k, \varkappa_k, r_{0,k} \}$. More details are given in the Appendix.

\section{Simulation results}
\subsection{Effective potential parameters from the optimization}
\begin{figure}[H]
\centering
\includegraphics[width=\textwidth]{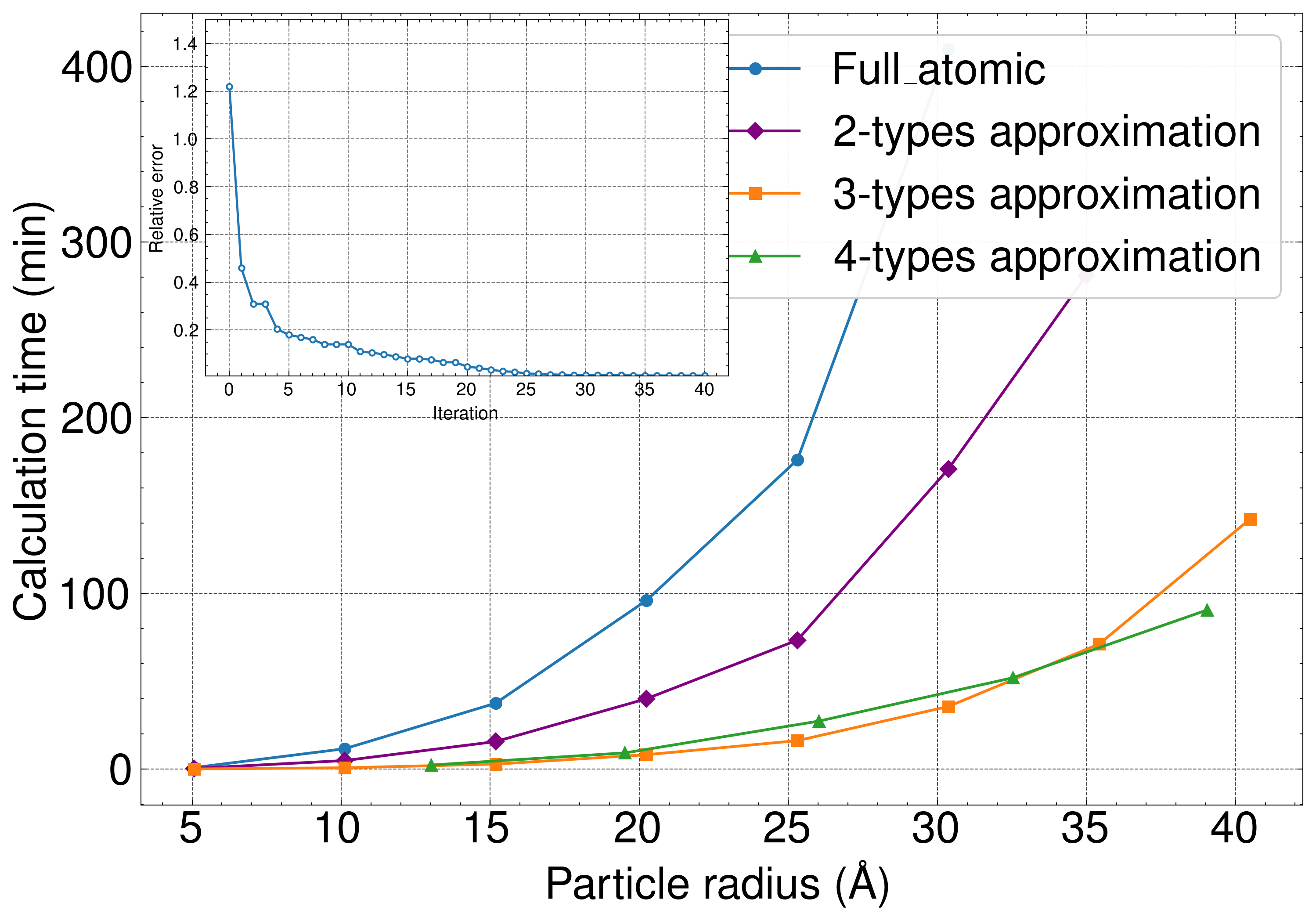}
    \caption{Main panel: The computation time dependence for inter-particle collisions,  on the particle size for the original atomistic particles (blue line) and multi-scale particles: with $N=2$ types of atoms/quasi-atoms (lilac line), $N=3$ types (orange line) and $N=4$ types  (green line). The original atomistic particles interact via standard Lennard-Jones potential corresponding to copper, while the potential parameters for the quasi-atoms are found by the optimization procedure. Inset: The dependence of the relative error for the target parameters, associated with the elastic constants of copper, on the iteration step of the optimization procedure for $N=4$ (see the main text). 
    }
    \label{fig:speedup}
\end{figure}

The optimization procedure of the previous section has been implemented for the multi-scale medium consisting of $N$ different types of atoms/quasi-atoms, namely, for $N=2$,  $N=3$ and $N=4$. We observe a  quick convergence of the relative error of target parameters, such that they  differ from the desired ones for less than $0.94\%$ after 30-40 iterations, see the Inset in  Fig. \ref{fig:speedup}.  This requires a rather limited amount of computation time, thus confirming the numerical efficiency of the approach. 
Note that the optimization procedure automatically tunes the  quasi-atom potential parameters to precisely match the target elastic  moduli. The specific parameters  chosen for the atomistic potential do affect the magnitude of the target constants, but  have no impact on the transferability or relative accuracy of the quasi-atom model. For the particular case of Lennard–Jones (LJ) potential, the appropriate 
quasi-atoms'  parameters may be found  analytically (see Appendix for detail). For more complex potentials, the analytical derivation becomes intractable, necessitating an application of the general numerical optimization.  Also note, that while we search here for the potential parameters to fit the elastic constants, the approach can be extended to a wider set of target quantities, including, for instance, viscous constants. We expect that the total number of the parameters to be fitted, is large enough to model many material properties.

\subsection{Particles collision: Atomistic versus multi-scale modelling }
\begin{figure}[H]
    \centering
    \begin{subfigure}[b]{0.48\textwidth}
        \centering
        \includegraphics[width=\textwidth]{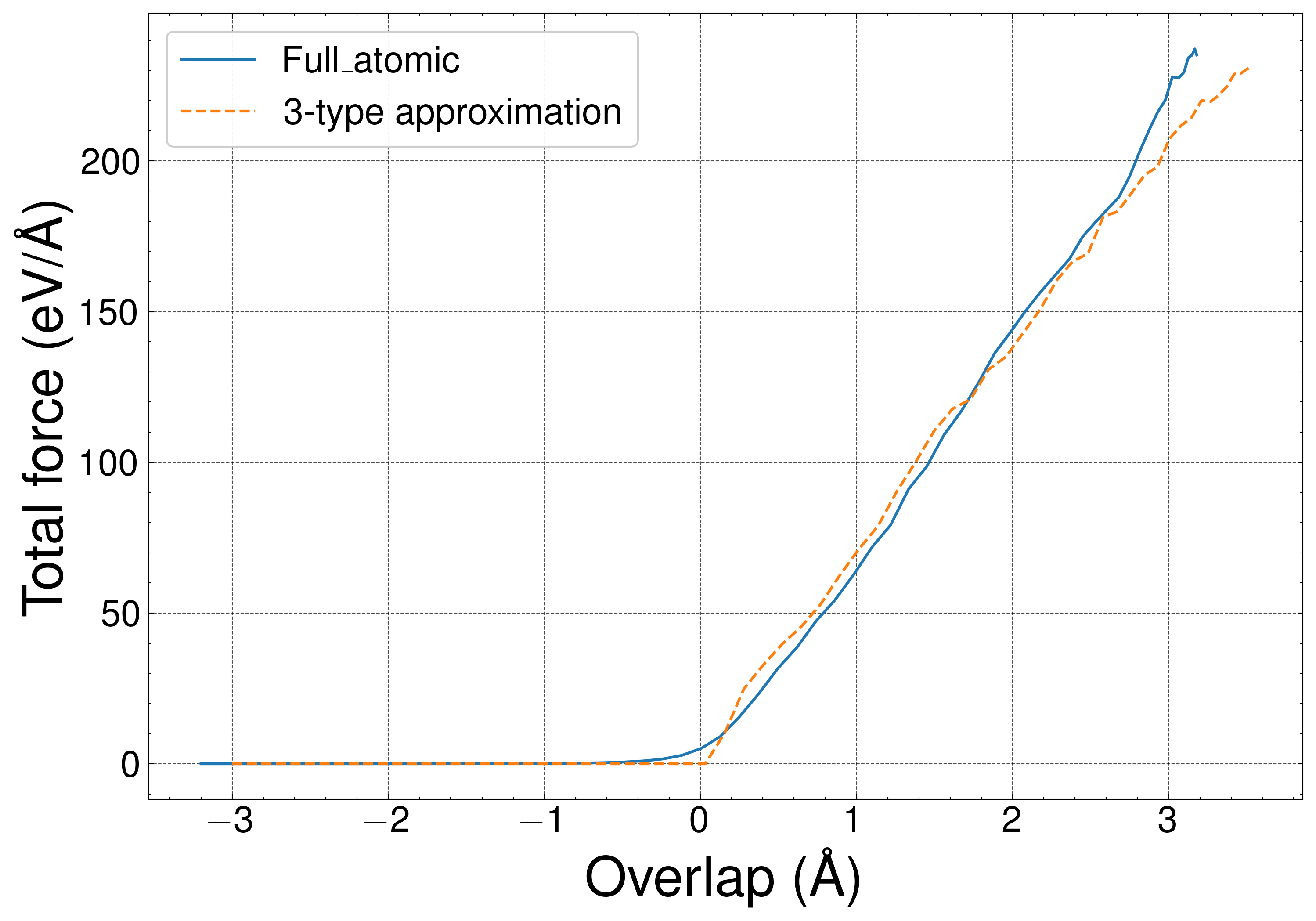}
        \caption{Non-attractive collision, $B=0$.}
        \label{fig:comparitive3}
    \end{subfigure}
    \hfill
    \begin{subfigure}[b]{0.48\textwidth}
        \centering
        \includegraphics[width=\textwidth]{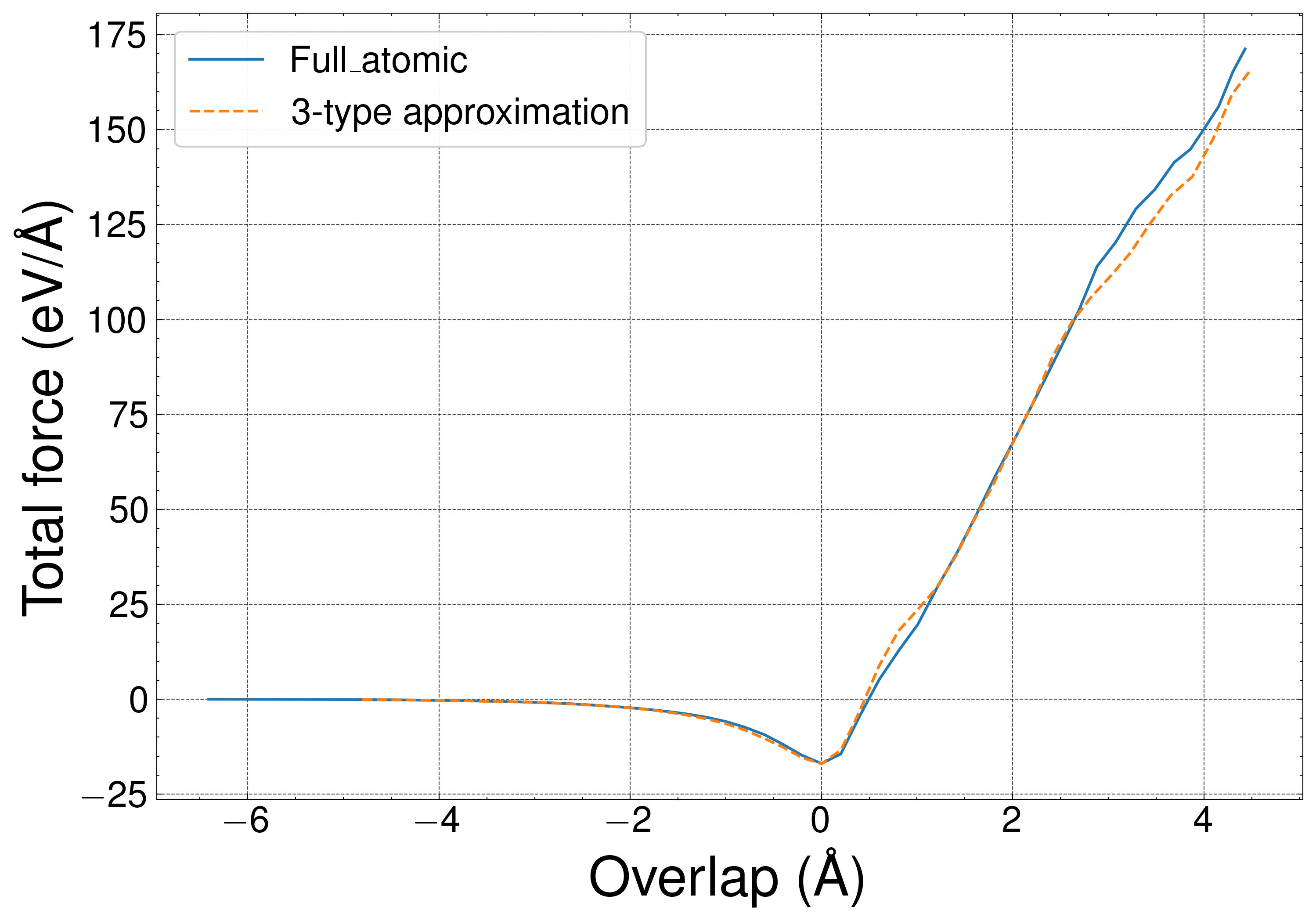}
        \caption{Collision with attractive forces, $B=0.3$.}
        \label{fig:comparitive4}
    \end{subfigure}
    \caption{Comparative analysis of the collision dynamics of  composite particles, comprised of atoms and quasi-atoms of three types and  completely atomistic particles (full-scale MD simulations). The inter-particle force is plotted as the function of the overlap, $\xi=R_1+R_2-r_{12}$.  The radii of the particles are $R_1=R_2=75\, \text{\AA}$, the impact velocity is $V=100\,m/s.$ The simulation time step is $\Delta t = 0.1\,\mathrm{fs}$ and cutoff distance for LJ potential is $r_{cut}=2.5\,\sigma$.}
    \label{fig:comparative_combined}
\end{figure}

To validate our multi-scale approach, we conducted comparative simulations of particle collisions, using both the composite  and full-atomic configurations. We consider spherical particles of radius $R=75 \,\text{\AA} $, colliding at the impact velocity of $V=100\,m/s$. Two distinct interaction regimes were analyzed -- purely repulsive and attractive collisions. For  the former case the value of  $B=0$ in Eq. \eqref{LJ} was used for the interaction potential between atoms, belonging to different colliding particles  (Fig.~\ref{fig:comparitive3}), for the latter one  we used $B=0.3$ (Fig. \ref{fig:comparitive4}).The temporal evolution and interaction forces demonstrate, in both cases,  a remarkable agreement, see  Figs.~\ref{fig:comparitive3} and \ref{fig:comparitive4}. These results confirm that our approach preserves the essential physics of the collision process, while significantly reducing computational complexity.

With the new tool at hand it is interesting to  analyze collisions of larger particles, with purely atomistic interactions at the point of  contact, when their size prohibits standard  full-atomic MD simulations. Hence, we explore the collision of spherical particles of micron size,  $R = 0.1 \mu m $.  Such composite objects correspond to particles comprising about $2\cdot 10^8$ atoms -- the number, hardly accessible for the current (conventional) computational facilities.

\begin{figure}[H]
    \centering
    \begin{subfigure}[b]{0.48\textwidth}
        \centering
        \includegraphics[width=\textwidth]{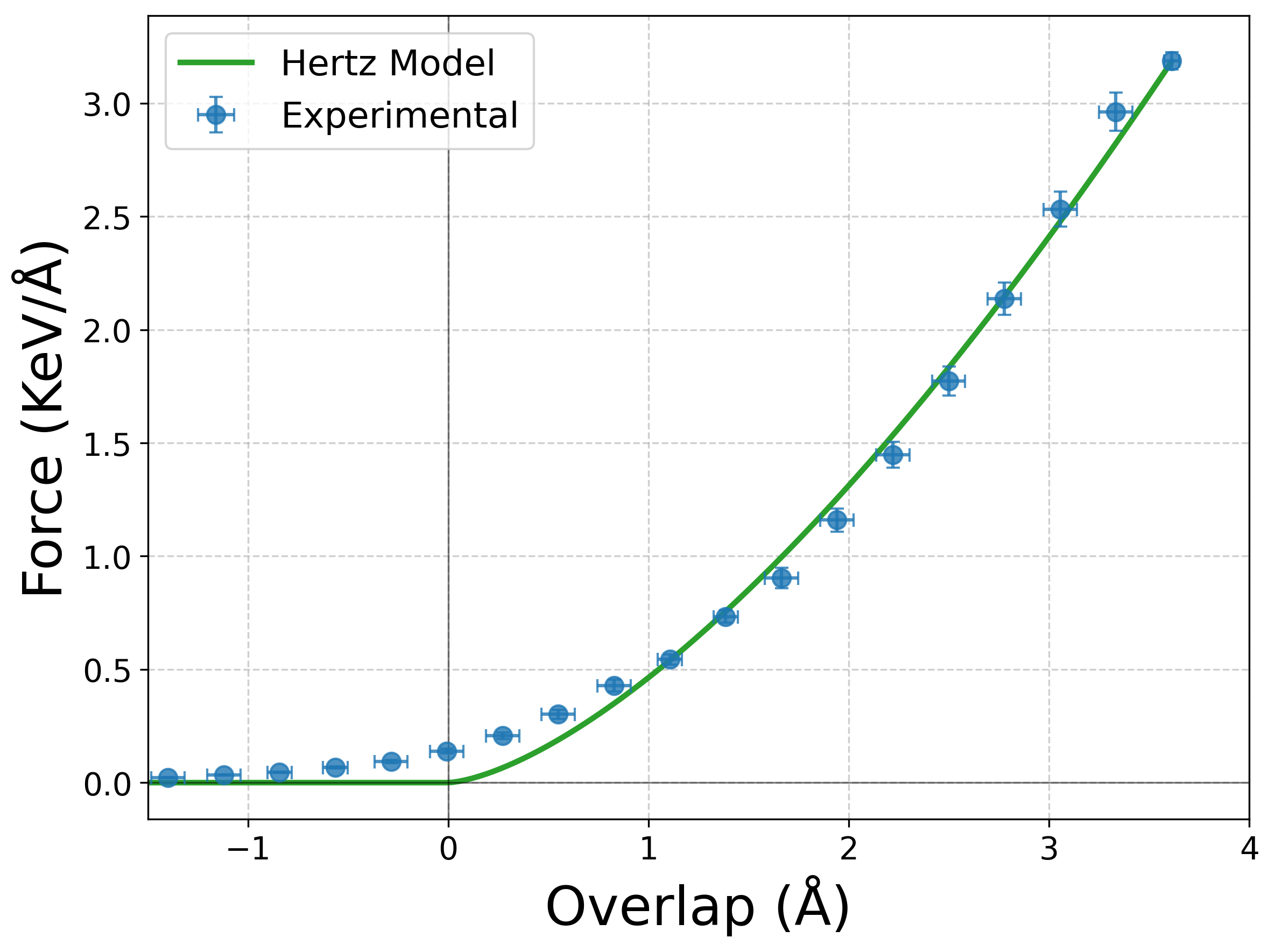}
        \caption{Non-attractive collision (LJ)}
        \label{fig:comparitive1}
    \end{subfigure}
    \hfill
    \begin{subfigure}[b]{0.5\textwidth}
        \centering
        \includegraphics[width=\textwidth]{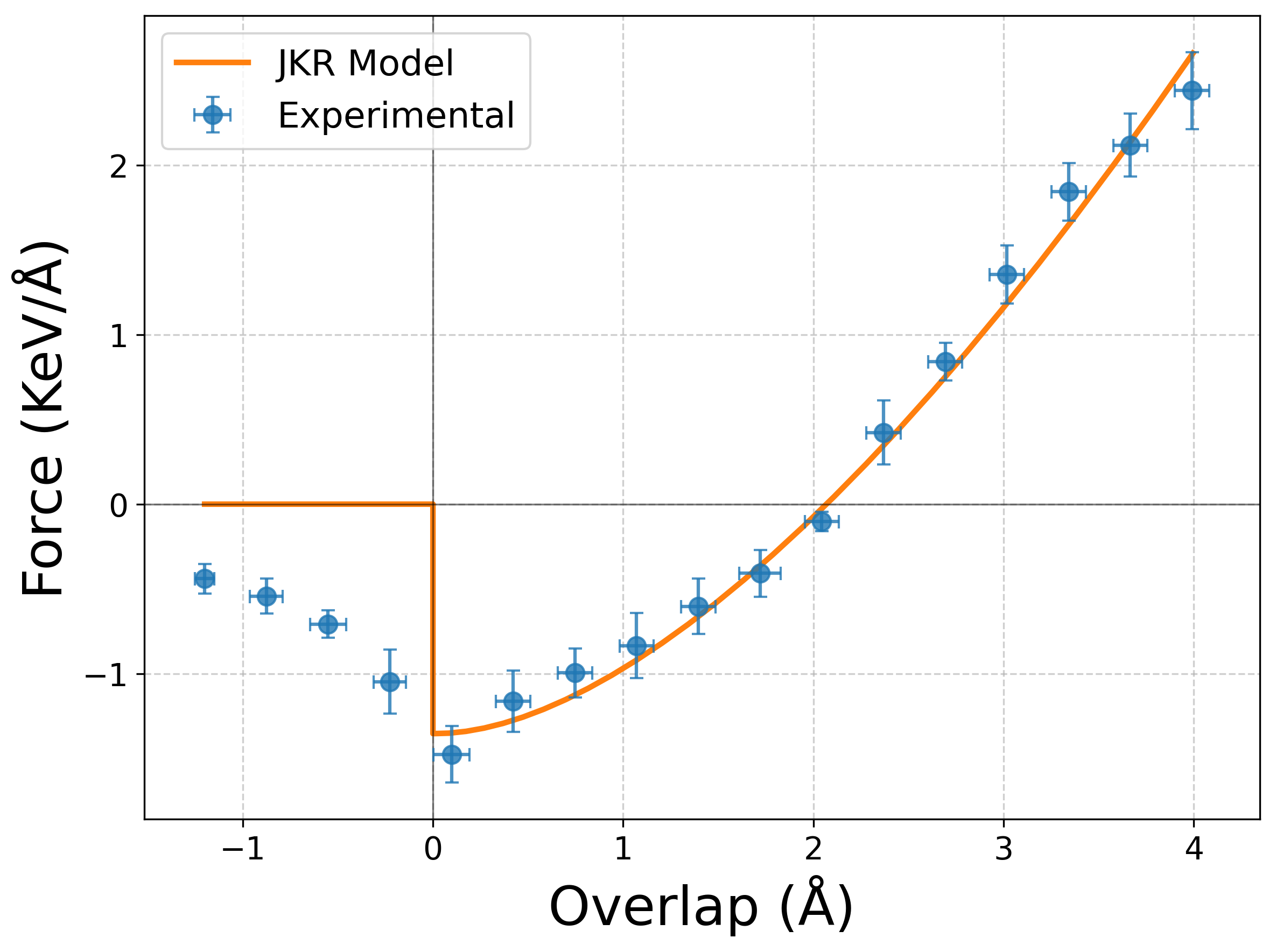}
        \caption{Attractive collision (LJ)}
        \label{fig:comparitive2}
    \end{subfigure} \\
    \begin{subfigure}[b]{0.49\textwidth}
        \centering
        \includegraphics[width=\textwidth]{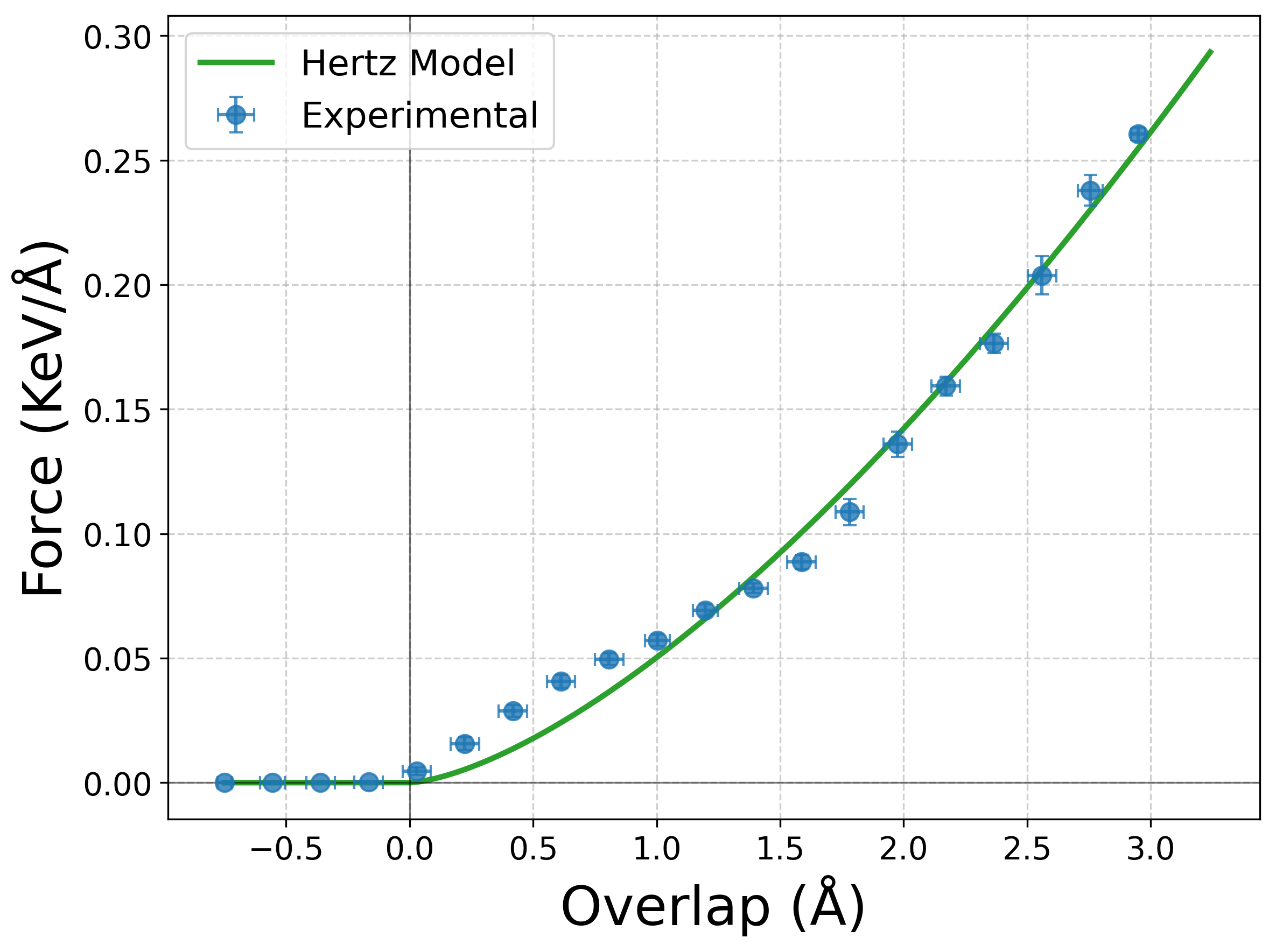}
        \caption{Non-attractive collision (Tersoff)}
        \label{fig:TersoffJKR1}
    \end{subfigure}
    \begin{subfigure}[b]{0.49\textwidth}
        \centering
        \includegraphics[width=\textwidth]{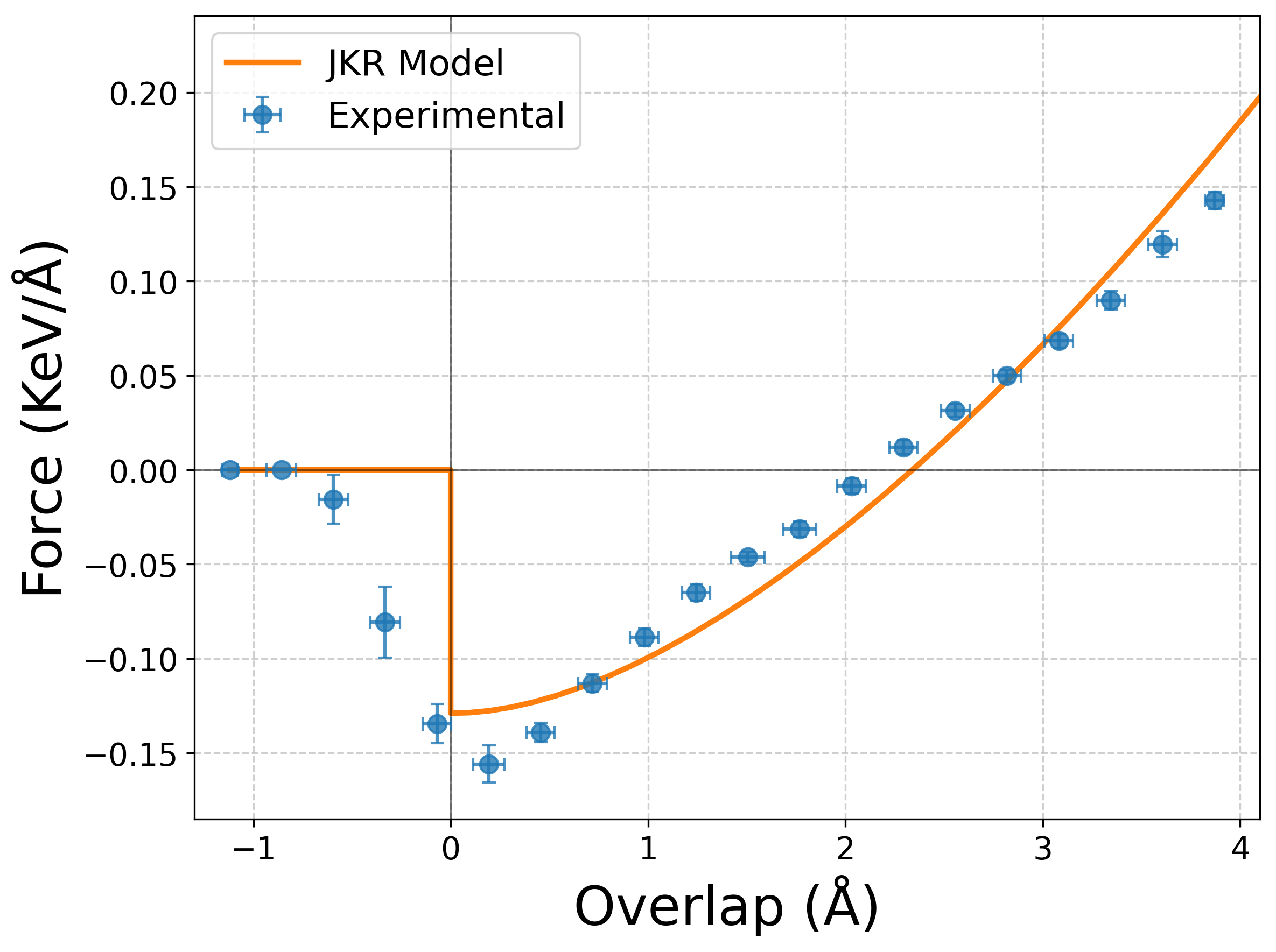}
        \caption{Attractive collision (Tersoff)}
        \label{fig:TersoffJKR2}
    \end{subfigure}
    \caption{Comparison of the collision dynamics of mesoscopic composite particles of radius $R=0.1\ \mu m$ with four types of atoms/quasi-atoms,  with the prediction of theoretical models --  the Hertz theory for non-adhesive contact (left) and JKR theory for the adhesive contact (right).  Two types of the inter-atomic potentials are demonstrated  -- the pairwise Lennard-Jones potential (upper panels) and multi-atomic Tersoff potential (bottom panels).      
The inter-particle force is plotted as the function of the overlap (see the text for detail). 
}
    \label{fig:theoretical_comparison}
\end{figure}

Again, we analyzed two interaction regimes -- purely repulsive  (Fig.~\ref{fig:comparitive1}) and collisions with an attraction (Fig.~\ref{fig:comparitive2}). The results were compared against classical contact theories - the Hertz  theory for non-adhesive contact \cite{Bril2007,Saitoh,Landau1965} and Johnson-Kendall-Roberts (JKR) theory for adhesive contact \cite{Bril2007,Saitoh}. Both of these  are macroscopic, static theories, based on the continuum elasticity theory. Therefore, to exclude the distortions due to dynamical effects, we performed quasi-static collisions. 

The interparticle force in these theories is expressed in terms of the overlap $\xi$-- the difference between the sum of particles radii $R_1$, $R_2$ and their inter-center distance $r_{12}=|{\bf r}_1-{\bf r}_2|$, that is, $\xi =R_1+R_2-r_{12}$. For the case of adhesive contact we consider, for simplicity,  only the first part of the collision, when the particles approach;  viz. we do not analyze the force hysteresis at the rebounce\footnote{The inter-particle force as well the overlap are expressed in the JKR theory in terms of the contact radius. Still, numerically,  the force may be expressed in terms of the overlap.}. 

The force-displacement curves obtained from our simulations show a nice overall agreement between the theoretical predictions and simulations for the positive overlap, that is, when particles are in a mechanical contact. A small discrepancy between the simulation results and Hertz theory at the overlap below $1\, \text{\AA}$ demonstrates the discrete, atomistic nature of the contact. Indeed, while the Hertz theory, based on elasticity theory of continuum, does not put a limit for a minimal overlap, the overlap for real atomistic contact is determined by the relative position of atomic layers of two particles. That is, a very small geometric overlap corresponding to a negligible force, computed by continuum theory, may cause a noticeable  inter-atomic repulsion detected in the simulations. 

As expected, significant deviations between simulations and the macroscopic JKR theory are  observed for the negative overlap, $\xi <0$, where the mechanical contact is lacking. Indeed,  JKR does not account for non-contact atomic interactions, although for $\xi <0$ the particles experience non-contact attractive van der Waals forces.  A small shift of the potential well of about $0.2\text{ \AA}$, observed in Fig.~\ref{fig:comparitive2}, may be attributed to the discreteness of atoms, implying a small ambiguity in defining the contact point. 

To further validate the generality of our approach and check its applicability beyond pairwise potentials, we performed a collision simulations of large silicon particles, interacting via the Tersoff potential \cite{tersoff1988new} with standard interaction parameters for silicon \cite{tersoff1988empirical} . This is a multi-atomic potential, used for materials with covalent bonds. We again consider two cases -- only repulsive collisions (with suppressed attractive forces) and attractive collisions, with the full potential. The results are compared with Hertz and JKR theory, see  Fig. \ref{fig:TersoffJKR1} and \ref{fig:TersoffJKR2}. Similarly as for the copper particles, the theory agrees reasonably well with simulations in the region of mechanical contact, $\xi>0$. For the attractive collision,  noticeable deviations   between the theory and simulations are observed for the  lack of contact, $\xi < 0$; naturally, the attractive van der Waals forces are not inclued in the JKR theory.

Plotting the theoretical curves in Fig. \ref{fig:theoretical_comparison}, we used the elastic constants obtained in our simulation. Although the adhesion coefficient of the JKR theory was treated as a fitting parameter (generally it may depend on a particle size), the according values, $\gamma= 0.3 \text{ J/m}^2$ for copper  and $\gamma = 0.05 \text{ J/m}^2$ for silicon particles are consistent with the range of adhesion values established in experimental and theoretical studies \cite{packham2003adhesion, pfaff2010coefficient, joy1990interface, kim2023high}.

Hence, an important conclusion follows from our simulations: The collision of mesoscopic (micron-size) particles may be adequately described by continuum  theories only for purely repulsive collisions. In the presence of attractive forces the continuum JKR theory should be supplemented by non-contact attractive van der Waals forces. 
Note that this conclusion is based on the micron-size particles simulations,  while maintaining the atomic-level accuracy in the critical (contact) region. 

\subsection{Computation efficiency of the multi-scale approach}
The developed multi-scale approach dramatically reduces the computation time when simulating large systems. Simultaneously, it preserves the modeling accuracy in  critical regions, where the atomic-scale simulations are  needed. For the parts  of the system,  remote from the critical areas,  one can successively increase the size of quasi-atoms up to the desired value. 

Obviously, the computation speed-up from the usage of multi-scale quasi-atoms crucially depends on the system size and number of types of the quasi-atoms. 
The larger the system, and number of quasi-atoms types,  the larger the computational time gain. This is illustrated in the main panel in Fig. \ref{fig:speedup}. As one can see from the figure, for the particle size of $30\, \text{\AA}$ the composite particles simulations are ten times faster than that for  the full-atomic simulations. For the size of $40\, \text{\AA}$ the simulation acceleration is overwhelming. 

As it also follows from Fig.~\ref{fig:speedup}, the computation time drastically decreases with increasing number of types of quasi-atoms: To model particles' collision with four  types of atoms/quasi-atoms is about four times faster, than with two types, already for the particles' radius $R=35 \,\text{\AA}$. Such an advantage quickly accrues with  the increasing particle size. 

One should however note, that the bigger the quasi-atoms, the lower the discretization  accuracy. This may lead to some mass deficiency, which is to be properly taken into account. Nevertheless it becomes negligible for large particles, see  the discussion in the Appendix.

\section{ Comparison of the  present method with other approaches}
\label{comparison}
There exist several multi-scale methods, addressing  bridging the atomistic and  continuum description; each of them, however, possesses significant limitations. 
The peridynamics approach, for instance, utilizes a non-local continuum formulation for modeling elastic behavior, see e.g. \cite{SILLING2000175}.  Yet, it  does not include an explicit atomic description; this reduces the modeling fidelity of the processes involving atomistic events (e.g., dislocation nucleation, crack initiation, etc.). 

The heterogeneous multiscale method (HMM) \cite{E2007_HMM_review} couples a macroscopic model with local molecular‐dynamics simulations to supply missing constitutive data, but it does not maintain a continuous atomistic simulation throughout the domain. This can limit HMM applicability for such phenomena as crack initiation and propagation,  where the atomic structure of the critical area evolves dynamically. Moreover, when the micro‐sampling region is displaced, or constitutive assumptions,  underlying the interpolation are violated, the accuracy of the method may substantially  degrade.

Coarse-grain (CG) methods remain widely used for fluids and soft matter \cite{Khan2022_CG_review}, and although  solid‐state CG schemes are available, they often struggle to capture full elastic anisotropy, fracture behavior or atomic‐scale mechanisms with high fidelity. 

\begin{figure}[H]
    \centering
    \includegraphics[width=\textwidth]{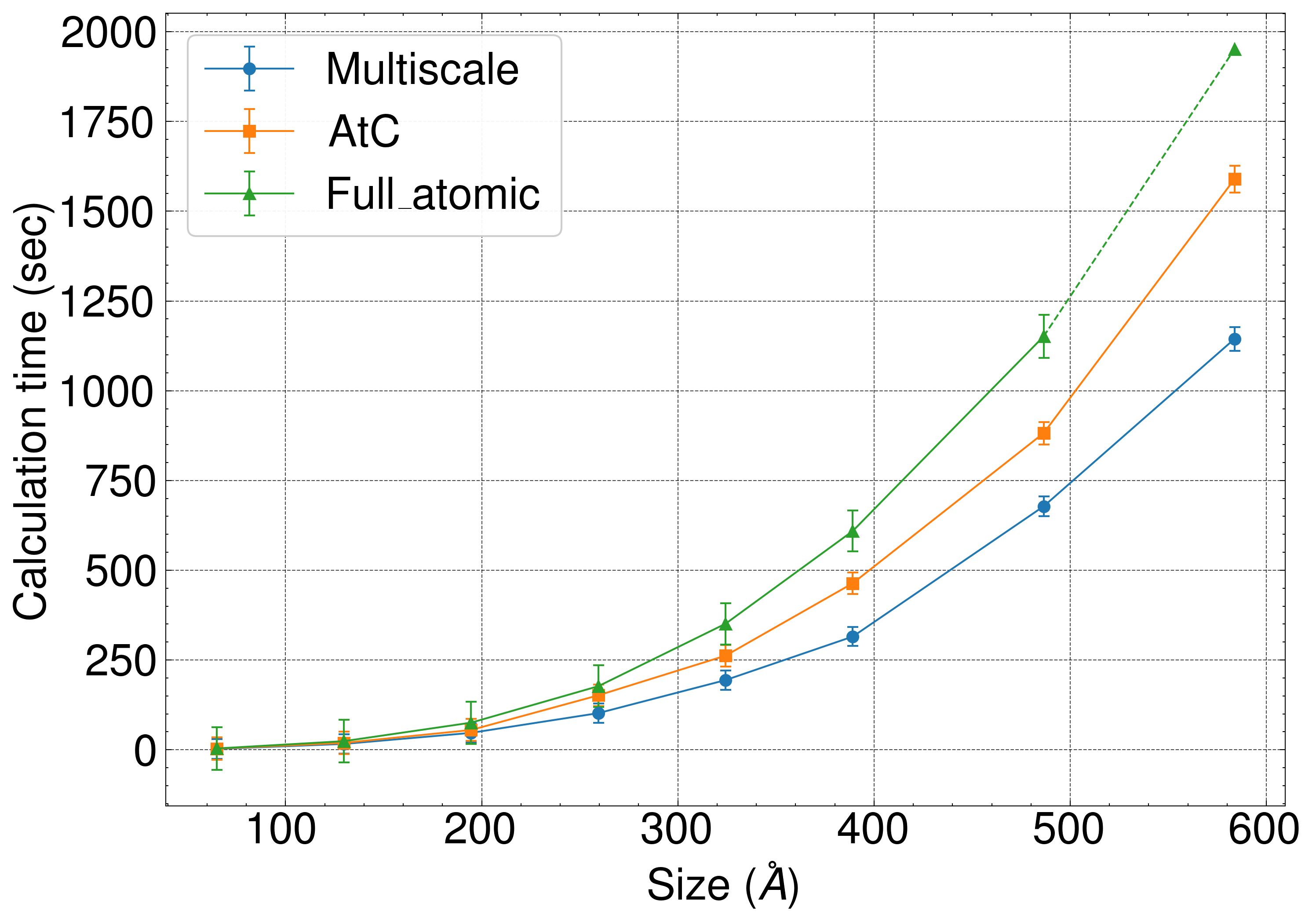}

\caption{Computation time for  equilibrium copper system at T=300 K,  as a function of the system size. See the text for details. 
}
    \label{fig:methods_comparison}
\end{figure}

There exist a couple of hybrid methods, e.g. Quasi-Continuum (QC) method \cite{MillerTadmor2002_QC_review, Ortner2008_QC_analysis}. Here we focus on  the AtC (Atoms-to-Continuum) method, which seems to be most close to our approach.   AtC is  available in LAMMPS as the USER-AtC package \cite{XuChen2018_AtC_review}. AtC uses blending functions to couple finite element meshes with atomic regions. Still, AtC remains challenging to use: it sufferes from the  solver overhead, complex setup, difficulty of construction of a seamless transition between the atomistic and continuum part. The latter follows from the lack of intrinsic relations between the inter-atomic potential parameters and the mesh size, which is chosen by trial and error. This leads to the loss of accuracy in the material modeling.  In practice, this means  that our hybrid scheme is faster, simpler to deploy, and more precise in the target metric, at least for the addressed material problems. That statement is supported by Fig.~\ref{fig:methods_comparison}, where we compare the computation time for particles' collision for full-atomic, AtC and our multi-scale method.

To compare the efficiency of different methods, independently of a particular  simulated process, we decide to model a system "without processes", that is, an equilibrium  system. Hence, we simulate copper at 300 K. We use the simulation box of varying size and (i) fill it completely with copper atoms,  (ii) fill a half of the box with copper atoms and half with  quasi-atoms of one type  -- twice larger than real atoms, and (iii) fill  a half with copper atoms and make a uniform finite-element mesh for the other half. For the item  (i) full-atomic simulations are performed, for the item (ii) -- multi-scale hybrid simulations and for (iii) -- AtC approach is applied. In this case we use $3\times 3\times 3 =27$ finite elements for the minimal box size of $50 \times 50 \times 50 \, \text{\AA}$, with the proportional increase of the number of elements with increasing box size. The goal was to simulate an equilibrium system over a fixed interval \({\cal T} = 100\,\mathrm{fs}\), using a time step \(\Delta t = 0.1\,\mathrm{fs}\) (1000 steps in total)  and compare the amount of computer time for all three methods. Note, that  during the equilibrium simulations one needs to solve Newton's equations of motion for atoms and quasi-atoms for the full-atomic and multi-scale method. At the same time, for the AtC approach one needs simultaneously solve  Newton's equations, make  the adjustments of properties of finite elements (in accordance with the positions of adjacent atoms)  and perform operations with the elements themselves. This requires a lot of computer  time, even for a relatively coarse mesh. We deliberately use in  our multi-scale approach only two types of atoms/quasi-atoms ($N=2$). This yields a significantly higher resolution of the multi-scale methods,  than that of AtC -- the number of mesh elements is 50 time less than the number of quasi-particles.   Yet the the multi-scale method is noticeably faster.  With the increasing $N$ (e.g. for $N=4$) such a superiority will be much more pronounced, see Fig.~\ref{fig:speedup}. In the Table \ref{tab:methods_booktabs} we present the assessment of the most popular multi-scale method. As it may be seen from the Fig.~\ref{fig:methods_comparison} and Table \ref{tab:methods_booktabs}, the proposed approach outperforms all the existing methods.

\begin{table}[H]

\centering

\small

\begin{tabular}{@{} l c c c c c @{}}

\toprule

\textbf{Method} & 

\shortstack{\textbf{Atomistic}\\\textbf{simulation}} & 

\shortstack{\textbf{Main}\\\textbf{phase}} &

\shortstack{\textbf{Open-source}\\\textbf{impl.}} &

\shortstack{\textbf{Comp.}\\\textbf{speed}} \\

\midrule

This work (Hybrid)    & \checkmark & Solids &  \checkmark & High \\

Quasi-Continuum (QC)  & \checkmark & Solids       & $\circ$    & Med. \\

AtC (LAMMPS)          & \checkmark & Solids        & \checkmark & Med. \\

Coarse-Grained / AdResS & $\circ$ & Liq./Gas      & $\circ$    & High \\

HMM                   & $\circ$    & All (macro)    & $\circ$    & Case-dep. \\

Peridynamics          & \ding{55}  & Solids      & \checkmark & Med. \\

\bottomrule

\end{tabular}

\caption{Compact comparison of hybrid and mesoscale methods. Symbols: \checkmark\,=\,yes; \ding{55}\,=\,no; $\circ$\,=\,partial/limited.}

\label{tab:methods_booktabs}

\end{table}

\section{Conclusions}
\label{sec:conclusions}

We have developed a novel hybrid multi-scale simulation approach, capable to accurately  describe both the atomic-scale interactions and dynamics of continuum media. The main idea of the method is to use simultaneously full-atomic simulations for the  critical regions, such as e.g. areas of particles' contacts, cracks' surfaces, etc. and continuum description of the rest of the systems using quasi-atoms. That is, the continuum part of the system is modelled as an effective medium comprised of quasi-atoms of successively increasing sizes with increasing distance from the critical regions. Depending on the system dimensions and the addressed problem, one can use as many types of quasi-atoms as needed. The parameters of the interaction potential between the quasi-atoms are found by the optimization procedure. The goal of this procedure is to find such values of the potential parameters, that yield the same properties of the effective medium as of the real, atomistic one. To make the optimization procedure computationally efficient, we employ a powerful tool of machine learning -- the so-called ”surrogate optimization”. It yields the accuracy better than 1\% for the elastic properties of the effective medium.   

The quasi-atoms concept is a very valuable feature of the new method, since it allows a straightforward application of standard software simulation packages; in our study we utilize LAMMPS. Thus, we have developed  a Python-based optimizer for obtaining the quasi-atoms potential parameters, as a flexible supplementary to LAMMPS.  Our Python–LAMMPS bridge enabled seamless coupling between optimization and simulation engines, allowing a robust derivation of parameters even for complex many-body potentials.

We validated our approach through detailed simulations of particle collisions. We explore the case of Lennard-Jones pairwise inter-atomic potential (to model copper particles) and multi-atomic Tersoff potential (to model silicon particles). 
Firstly, we analyze the impact of particles of relatively small size of $R=75\,\text{\AA}$, which allows full-atomic simulations. We compare the collision dynamics of real, atomic-based particles, and composite particles, built up of three different types of quasi-atoms. We observe an excellent agreement between these two dynamics, which confirms the validity and  accuracy of the new simulation method.     

Next, we apply our approach to explore the impact of particles at the bmicron scale,  $R = 0.1 \mu m $ -- the regime,  commonly  inaccessible to fully atomistic methods. The simulation  results are compared with the predictions of continuum  macroscopic theories --  Hertz theory for non-adhesive contacts and JKR theory for  adhesive interactions. Our results prove the validity of the macroscopic Hertz theory and JKR for the regime of a full mechanical contact. Nevertheless, we observe that JKR theory should be supplemented with non-contact attractive van der Waals forces, essential for micron/submicron  particles. 

We demonstrate that our approach leads to a drastic reduction of the computation time already at the micron level and manifests the increasing superiority with increasing system size.  The reported results confirm that the new method  successfully preserves key mechanical and dynamical properties of a system, simultaneously offering substantial computational savings. This paves a way for an accurate modeling of many mesoscale phenomena, such as fracture, sintering, and self-assembly, without compromising atomic resolution in critical regions. Moreover, the method may be straightforwardly extended for macroscopic systems.

 We compare our method with other multi-scale methods and perform   comparative simulations with the most close AtC method. We demonstrate that the proposed hybrid approach  is faster, simpler to deploy, and more precise in the target
metric, than other methods, at least for the addressed material problems. 

Finally, we note that the core methodology of the new method relies on a general black-box optimization of the interaction parameters. That is, the parameters of the interaction potential  of quasi-atoms,  comprising the effective medium; these are optimized to match  macroscopic constants of real continuum.  Here we fit the elastic constants, but many other continuum characteristics may be  fitted as well. Currently, we do not see any theoretical limitations to apply the new method for other non-equilibrium processes, 
such as thermal transport, dynamic fracture propagation, etc., as long as the main part of the system may be described as continuum. In particular, this would imply, that only acoustic phonons are important for a  system dynamics. More precisely, the maximal size of quasi-atoms, $\sigma_{\rm max}$,  should be much smaller than the minimal scale of a process of interest, ${\cal L}$, viz. $\sigma_{\rm max} \ll {\cal L}$. For instance, when studying a propagation, of ultrasound, which  maximal frequency $f$ may reach Gigahertz, $f \sim 10^9 \,s^{-1}$, the minimal size, needed to be resolved, is the sound wave length, ${\cal L}\sim \lambda \sim c/f$ ($c $ is the speed of sound).  Using $c=4.6\, 10^3\, m/s$ for copper, we obtain ${\cal L} \sim 4.6\, \mu m$, which is about 20 000 larger than the size of copper atom. Such an estimate demonstrates that the quasi-atom approach, in practice,  does not have limitations in modeling continuum media in terms of its scales.

Future studies will focus on extending the methodology to a wider range of materials and geometries and more complex and realistic interaction potentials, including many-body force fields, such as e.g. EAM (see e.g. \cite{PhysRevB.29.6443}). Additionally, more optimization objectives -- to match solid viscosity, thermal conductivity, fracture toughness, etc. will be incorporated into the optimization procedure. This will enable a broader class of multi-scale simulations. Furthermore, the optimization procedure itself, may be recast into  the form of reinforcement learning problem in the field of artificial intelligence \cite{barto2021reinforcement}, specifically in the form of   multi-armed bandit problem  \cite{lattimore2020bandit}. This will allow to exploit a number of computationally powerful tools \cite{shakya2023reinforcement} to significantly increase the optimization efficiency. 

\appendix
\section{Optimization procedure}
\label{Opt_appen}

\subsection{Construction of the surrogate loss function}

Conceptually the optimization procedure  coincides with Machine Learning (ML). Indeed, a loss function in the ML characterizes a deviation of some quantity (e.g. a number encoding features in the ML) from the target one. This quantity depends on a system parameters of interest and the loss function is to be iteratively minimized, with respect to these parameters,  to achieve its global minimum. To search for the global minimum one needs to combine both --  exploration and exploitation algorithms. The former is used to explore all needed regions of the parametric space, the latter -- to find a minimum of the loss function in the specific  space region. Hence, the optimization is rather computationally expensive, especially if MD simulations for each iteration step are required. This motivates an application of surrogate loss function discussed below for the black-box optimization. The main goal of constructing such a function is to reduce the number of full-scale MD simulations, during the optimization procedure. Here we use a computationally efficient method of radial basis functions (RBF) 
 \cite{regis2007}. Denote by ${\bf x}$ the extended parametric set, 
$$
{\bf x} =\left(\sigma_1, \epsilon_1, \varkappa_1, r_{0,1}, \ldots, 
\sigma_N, \epsilon_N, \varkappa_N, r_{0,N}\right) \equiv \{\sigma_k, \epsilon_k, \varkappa_k, r_{0,k} \}, 
$$ 
comprising all parameters describing the composite medium, that is, ${\bf x}$ is a vector with  $4 \,(N-1)$ components (recall, that $k=1,2 \ldots, N-1$). 

Generate  now   uniformly $n$ random points ${\bf x}_i$  with $i=1,2\ldots, n$ in the appropriate part of the parametric space, that is, $n$ random vectors ${\bf x}_i$. The respective area in the parametric space is chosen, based on the approximate theoretical estimates. 

For instance, for \(N\) quasi‐atom types (for the notation brevity we use $N$, and  not $N-1$ below)  with a linear size scaling factor of two, the estimated Lennard–Jones parameter for type \(k\) reads (see the respective derivation below, in ~\ref{app:lj_derivation}):
\begin{equation}
  \hat\epsilon_i = 2^{3(k-1)}\,\epsilon_1,
  \quad k = 1,2,\dots,N,
  \label{eq:epsilon_estimate}
\end{equation}
where \(\epsilon_1\) is the atomic LJ constant (it is fixed and hence is not optimized).  This allows to perform the optimization of  each \(\epsilon_k\) within a relatively  small interval:
\begin{equation}
  \epsilon_k \in \bigl[\hat\epsilon_k - \Delta_k,\;\hat\epsilon_k + \Delta_k\bigr],
  \quad  \quad  k = 2,\dots,N.
  \label{eq:epsilon_region}
\end{equation}

Using $n$ points ${\bf x}_i$ we construct the surrogate loss function: 
\begin{equation}
\label{SL}
\hat{L}({\bf x}, {\bf \theta}) = \sum_{i=1}^n w_i\phi(\|{\bf x}-{\bf x}_i\|) + p({\bf x}), 
\end{equation}
where it is convenient to use the thin-spline function, 
\begin{equation}
\phi(||{\bf x} ||) = ||{\bf x} ||^2\log(||{\bf x} ||),
\end{equation}
and $p({\bf x})$ is the $m$-order polynomial with the constants $\lambda_{i_{1,1},i_{2,1}, \ldots, i_{4,N}}$:
\begin{equation}
\label{polyn}
p({\bf x}) =\underbrace{\sum_{i_{1,1}=0}^m \ldots \sum_{i_{4,N}=0}^m}_{i_{1,1}+i_{2,1}+\ldots +i_{4,N} \leq m}  \lambda_{i_{1,1},i_{2,1}, \ldots, i_{4,N}} \, \sigma_1^{i_{1,1}} \epsilon_1^{i_{2,1}} \varkappa_1^{i_{3,1}}  r_{0,1}^{i_{4,1}} \ldots 
\sigma_N^{i_{1,N}} \epsilon_N^{i_{2,N}}  \varkappa_N^{i_{3,N}} r_{0,N}^{i_{4,N}}.
\end{equation}
With the above notations  the vector $\theta$ is the concatenation of the  sets  $\{w_i\} $ and $\{ \lambda_{i_{1,1},i_{2,1}, \ldots, i_{4,N}} \}$, that is,  
$$
\theta=\left(w_1, \ldots w_n,  \lambda_{0, 0, 0,0 }, \ldots  \lambda_{m, 0, 0,0 }\right). 
$$
To illustrate the meaning of $p({\bf x})$ we write it for the simplified case of one type of quasi-atoms, when only parameters $\epsilon_2$ and $\varkappa_2$ are to be optimized (keeping $\sigma_2$ and $r_{0,2}$ constant). In this case the second order polynomial ($m=2$) has the form: 
$$
p({\bf x}) =\lambda_{2,0} \epsilon_{2}^2 + \lambda_{1,1} \epsilon_{2} \varkappa_2 +\lambda_{0,2} \varkappa_2^2+ \lambda_{1,0} \epsilon_{2} +\lambda_{0,1} \epsilon_{2} \varkappa_2 +\lambda_{0,0} , 
$$
so that the vector $\theta$ reads $\theta=\left(w_1, \ldots w_n,  \lambda_{0, 0}, \ldots  \lambda_{2, 0}\right)$. 

Now we compute the actual loss function in $n$ points ${\bf x}_i$,

\begin{equation}
  \label{eq:L_pq}
  L\bigl(\mathbf{x}_i\bigr)
  = \frac{1}{6}
  \sum_{p,q \,\in\,\{x,y,z\}}
    \frac{\bigl|\mathcal{E}_{\mathrm{target}}^{pq}
      - \mathcal{E}_{\mathrm{sim}}^{pq}(\mathbf{x}_i)\bigr|}
         {\mathcal{E}_{\mathrm{target}}^{pq}} \,,
\end{equation}
and require that it coincides with the surrogate one, $\hat{L}({\bf x}_i) = L({\bf x}_i)$. This yields the system of algebraic equations for the coefficients   \(w_i\) and the polynomial coefficients $\lambda_{i_{1,1},i_{2,1}, \ldots, i_{4,N}}$. Unfortunately, such a straightforward approach does not lead to an optimal approximation. A special analysis has been performed in Ref.  \cite{regis2007}, which shown that an additional constrains are to be imposed on the polynomial part of the surrogate loss function; this helps to make the approximation significantly smoother. Referring for detail to the original paper \cite{regis2007}, we present here the final set of equations, written in the matrix form: 

\begin{equation}
\label{main}
\begin{pmatrix} \Phi & P \\ P^T & 0 \end{pmatrix} \begin{pmatrix} W \\ \Lambda \end{pmatrix} = \begin{pmatrix} L \\ 0 \end{pmatrix}.
\end{equation}
Here $n\times n$ matrix $\Phi$ is defined as \(\Phi_{ij} = \phi(\|x_i-x_j\|)\). The matrix $P$ has $n$ rows and the number of columns associated with the number of terms in the polynomial in Eq. \eqref{polyn}; each element of this matrix in $i$th row and $j$-th column   is determined by the value of the $j$-th term of the polynomial \eqref{polyn},  taken at the point ${\bf x}_i$ without the respective constant coefficient. $P^T$ is the transposition of the matrix $P$. Respectively, $W$ is the $n$-dimensional vector $(w_1, \ldots w_n)$, while vector $\Lambda$ contains all the coefficients $\lambda_{i_{1,1},i_{2,1}, \ldots, i_{4,N}}$. In the right hand side of Eq. \eqref{main} $L$ is the $n$-dimensional vector with the components $\left( L({\bf x}_1), \dots,L({\bf x}_n) \right)$.

In practical computations we did not optimize such parameters as $\sigma_k$ and $r_{0,k}$, that is, we use the reduced vector ${\bf x}=\left(\epsilon_1, \varkappa_1, \ldots, \epsilon_N, \varkappa_N \right)$, while determining the length parameters $\sigma_k$ and $r_{0,k}$ from the geometrical lattice scaling.  We observe that such a reduction of the set of optimizing parameters keeps an acceptable accuracy of the approach,  significantly reducing the computation time.

To enhance robustness, we implement online feature scaling, transforming the input space ($x=(\{\epsilon_k, \varkappa_k\})$ in our case) to the interval \([-1,1]\) and the target values to the interval \([0,1]\):

\begin{equation}
\tilde{x} = 2\frac{x - x_{min}}{x_{max} - x_{min}} - 1, \quad  \quad \tilde{L} = \frac{L - L_{min}}{L_{max} - L_{min}}
\end{equation}
\\
\subsection{Surrogate function minimization}
After solving the equations \eqref{main} we obtain an explicit form of the surrogate loss function $\hat{L}({\bf x})$. Now, instead of minimization of the actual loss function ${L}({\bf x})$, which requires costly full-scale simulations, we minimize the surrogate loss function $\hat{L}({\bf x})$, assuming that the results of the minimization of $\hat{L}({\bf x})$ would be very close to this of ${L}({\bf x})$ (recall that these two functions are equal in the points ${\bf x}_i$). Hence we search for the optimal ${\bf x}$ through the following optimization:
\begin{equation}
\min_{x} \left[\hat{L}({\bf x}) - \alpha \min_{i} (\|{\bf x}-{\bf x}_i\|)\right], 
\end{equation}
where the additional term with \(\alpha\) stimulates the exploration and encourages  sampling in unexplored regions, thus, preventing a stack in a local minimum of $L({\bf x})$, see  Ref.\cite{regis2007}. The basin-hopping algorithm is employed to solve this optimization problem, which can be viewed as a Monte Carlo minimization that samples the transformed "energy landscape", see e.g. Ref. \cite{wales1997global}:

\begin{equation}
\tilde{E}(x) = \hat{L}(x) - \alpha \min_{i} (\|x-x_i\|)
\end{equation}

\subsection{Speed up by the parallel implementation}
We also applied the following trick, which allows to significantly speed up the optimization. Namely, we simultaneously explore a number of different parameters $\alpha$ from a wide interval. In practice, we run in parallel  \(K\) processes with the following different parameters $\alpha_l$:

\begin{equation}
\alpha_l = \alpha_{max}\left(1 - \frac{l}{K-1}\right), \quad l = 0,\ldots,K-1.
\end{equation}
This allows to find simultaneously $K$ best values  ${\bf x}_l$ values for each $\alpha_l$, and compute the additional $K$  new values of the real loss function $L({\bf x}_l)$.  These $K$ values of $L({\bf x}_l)$ are then used to extend the set of equations \eqref{main}, that is, the vector $L$ becomes now by $K$ terms longer. Solving then the extended set of equations yields an improved approximation for the surrogate loss function $\hat{L}$. The parallel optimization procedure can be summarized as:

\begin{align}
{\bf x}_{k,t+1} &= \arg\min_x \left[\hat{L}_{k,t}(x) - \alpha_k \min_{i} (\|{\bf x} - {\bf x}_i\|)\right] \\
L_{k,t+1} &= \hat{L}({\bf x}_{k,t+1})
\end{align}
where $\hat{L}({\bf x}_{k,t})$  is the surrogate loss function at $t$-th iteration.

\begin{figure}[H]
\centering
    \begin{subfigure}[t]{0.48\textwidth} 
       \includegraphics[width=\textwidth]{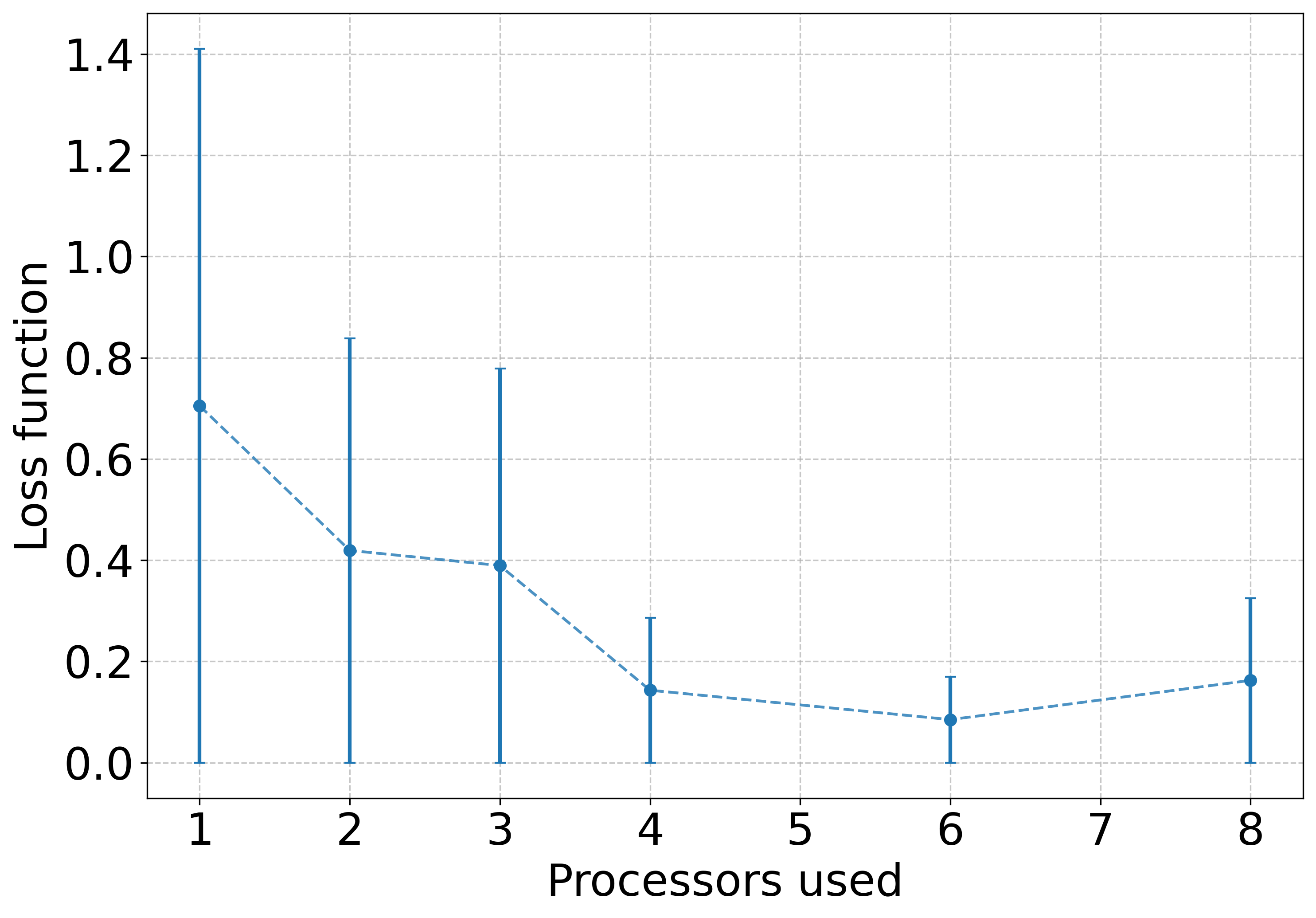} 
        \caption{The dependence of the  loss function at the beginning of the optimization procedure on the number of processors.}
        \label{fig:efficiency}
    \end{subfigure}
    \hfill 
    \begin{subfigure}[t]{0.48\textwidth} 
       \includegraphics[width=\textwidth]{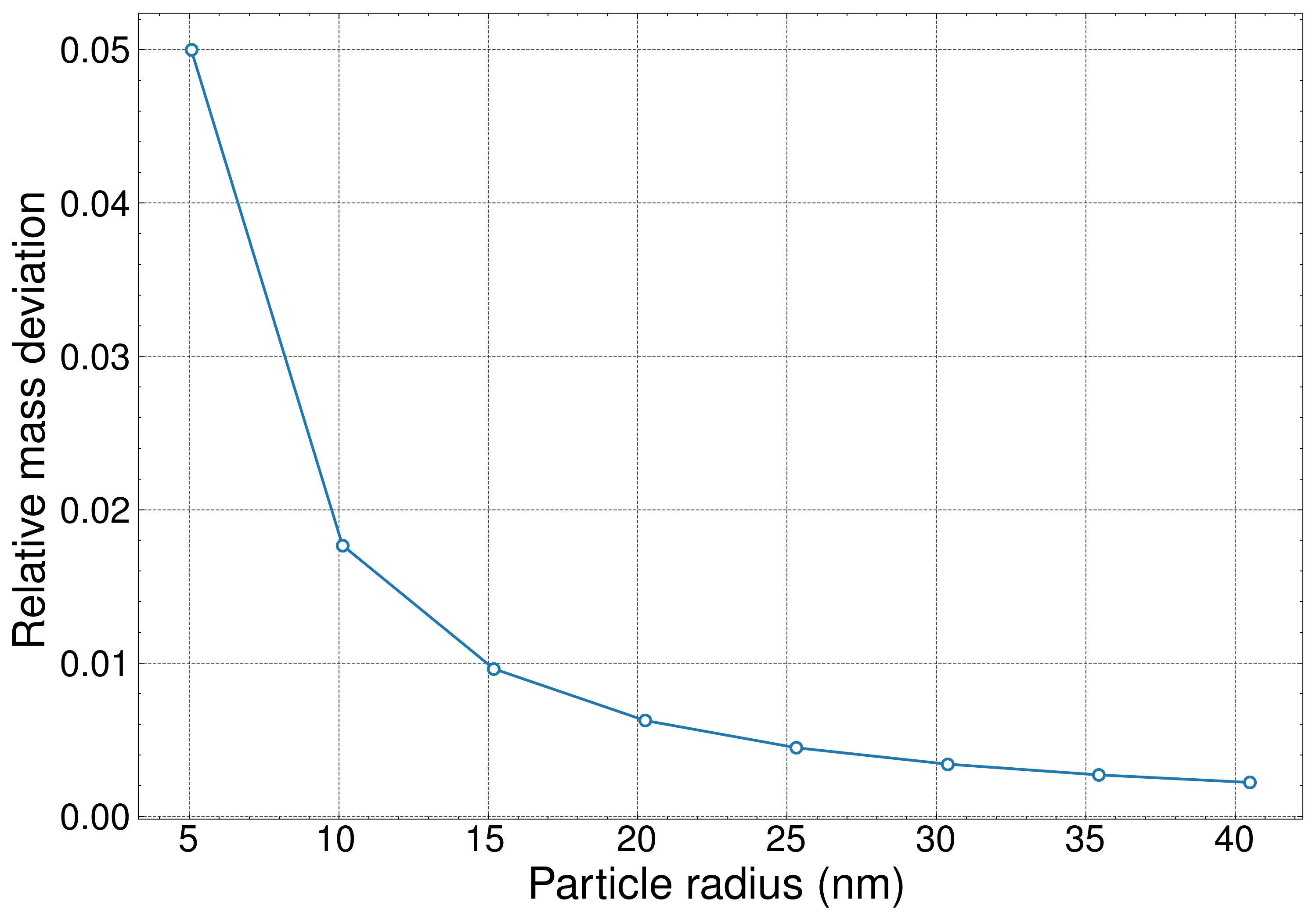} 
        \caption{Relative mass deficiency $\Delta m$ as a function of particle radius $R$ for composite particles with 3-types of atoms/quasi-atoms. }
        
        \label{fig:mass_def}
    \end{subfigure}
\caption{(a) The dependence of the loss function, $L$, for a fixed number of points, $n=24$, at the beginning of the optimization procedure  on the number of processors (see Sec. 2.4 for more detail) and  (b) relative mass deficiency $\Delta m$ as a function of particle radius $R$ for composite particles.}
         \label{FigApp}
\end{figure}

Parallelization efficiency in our scheme depends on several factors, such as the problem dimensionality, the computational cost of each iteration, and the choice of the parameter  $\alpha_{max}$. Since  our problem is computationally expensive, each iteration’s runtime is much greater than the data-transfer time; consequently, as the problem dimensionality increases (in our case, the number of potential parameters to optimize), the parallel efficiency approaches unity. It is also important to note that LAMMPS already provides CPU parallelization: repurposing those CPU threads to run additional exploration workers may therefore be questionable, although doing so is clearly advantageous for GPU runs. Thus, the central question is how many points one needs to compute.  We investigate, whether computing a batch of exploratory points, before updating the weights outperforms computing points sequentially with a separate update after each point. To assess an optimal number of processors, we analyze the loss function $L$ for a fixed number of points, $n=24$,  defined in Sec. 2.4, at the beginning of the optimization procedure, where $L$ starts to converge. The resulting dependence is shown in Fig. \ref{fig:efficiency}. As one can see from the figure, the increasing number of processors leads initially   to a more accurate and stable result, since the exploration is enhanced. However, at some point  the optimizer may have a lack of exploitation as the computer power becomes too much biased to the exploration.

\subsection{Regularization}
To control the overfitting, we add the L2 regularization to  coefficients $W=(w_1,\ldots, w_n)$, modifying the matrix equation to the following one:

\begin{equation}
\begin{pmatrix} \Phi + \gamma I & P \\ P^T & 0 \end{pmatrix} \begin{pmatrix} W \\ \Lambda \end{pmatrix} = \begin{pmatrix} L \\ 0 \end{pmatrix},
\end{equation}
where \(\gamma \) is the regularization parameter. This leads to the regularized optimization problem:

\begin{equation}
\min_{W,\gamma} \|BW - Y\|^2 + \gamma\|W\|^2
\end{equation}
where \(B = [\Phi \quad P]\) and \(Y = [L \quad 0]^T\).

\subsection{Analytical Derivation of Lennard–Jones Parameter Scaling}
\label{app:lj_derivation}

In this section we derive the scaling laws for the Lennard–Jones (LJ) parameters $\sigma$ and $\epsilon$ when a coarse‐graining  increases the linear size of the simulation cell by a factor of $K$.  We  show how this scaling preserves both the bulk potential energy density and the per‐atom stress used to compute elastic moduli in LAMMPS.

The derivation relies on two fundamental physical requirements: geometric consistency and thermodynamic conservation. Geometrically, when the lattice is coarsened by a linear factor $K$, the equilibrium spacing $\sigma$ must scale linearly to maintain the relative structural arrangement. Thermodynamically, to ensure the coarse-grained material exhibits the same bulk stiffness and cohesion as the atomic solid, the energy density (energy per unit volume) must remain invariant. These two constraints uniquely determine how the energy parameter $\epsilon$ must be rescaled.

Let $r_{ij}$ be the distance between two atoms $i,j$.  Under a coarse‐graining that doubles every linear dimension ($K=2$), distances scale as
\begin{equation}
  r_{ij}' \;=\; K\,r_{ij}.
\end{equation}
Since $\sigma$ defines the equilibrium distance at which the LJ potential reaches its minimum, it must scale identically,
\begin{equation}
  \sigma' \;=\; K\,\sigma.
\end{equation}
This ensures that the location of the potential minimum correctly reflects the enlarged coarse‐grained spacing.

The LJ potential between two particles reads
\begin{equation}
  V_{LJ}(r) \;=\; 4\,\epsilon 
  \Bigl[\bigl(\tfrac{\sigma}{r}\bigr)^{12} \;-\;\bigl(\tfrac{\sigma}{r}\bigr)^{6}\Bigr].
\end{equation}
In the coarse‐grained “quasi-atom” approach, one quasi-atom represents $K^{3}$ original atoms, so the total well‐depth energy contributed by that group is $L^{3}\epsilon$ when $r\!=\!\sigma$.  To preserve the same energy density (energy per coarse‐grain volume), we require
\begin{equation}
  K^{3}\,\epsilon \;=\;\epsilon'
  \quad\Longrightarrow\quad 
  \epsilon' \;=\; K^{3}\,\epsilon.
\end{equation}
For $K=2$, this gives $\epsilon'=8\,\epsilon$, similar to the eight‐fold increase in mass.

In LAMMPS the per‐atom stress tensor is computed (in pressure–volume units) via the {\tt "compute stress/atom"} command, summing kinetic and virial contributions as
\begin{equation}
  S_{\alpha\beta}^{(i)}
  \;=\;
  -\,m_i v_{i,\alpha} v_{i,\beta}
  \;-\;\tfrac12
  \sum_{j\neq i}
    \bigl(r_{ij,\alpha} F_{ij,\beta} + r_{ij,\beta} F_{ij,\alpha}\bigr),
\end{equation}
then, the macroscopic stress $P_{\alpha\beta}$ is obtained summing over the atoms $i$ and dividing by total volume.  Since $r_{ij}$ scales as $L$ and forces $F_{ij}\!=\!-\partial V/\partial r\propto\epsilon/\sigma$, one finds
\[
  r'\,F' \;\propto\; K\,r \;\times\;(K^{3}\epsilon)/(K\,\sigma)\;=\;K^{3}\,(rF)\,.
\]
Dividing by the new volume~$K^{3}\,V$ leaves the stress invariant, ensuring that the computed Young’s modulus (from the slope of stress–strain) remains unchanged under the coarse‐graining rules above.

Physically, doubling the linear dimension of each coarse‐grain cell both multiplies the number of atoms by eight and increases the spacing between equilibrium positions by a factor of two.  Scaling $\sigma\mapsto2\sigma$ preserves the geometric arrangement, while scaling $\epsilon\mapsto8\epsilon$ preserves the energy density (the total cohesive energy) and the virial contributions that underpin elasticity.  Together, these scalings guarantee that both the per‐atom energy density and the per‐atom stress (hence bulk elastic moduli) remain consistent between the fine‐grained and coarse‐grained descriptions.
\subsection{Other potentials}
Similar approach may be applied to other analytical potentials. In this section we show this the for three-body Tersoff potential.

The total potential energy reads, 
\begin{equation}
E \;=\; \frac{1}{2}\sum_i \sum_{j\neq i} V_{ij},
\end{equation}
where the the quantity 
\begin{equation}
\label{eq:Vij}
V_{ij} \;=\; f_C(r_{ij}+\delta)\,\Big[\,f_R(r_{ij}+\delta) + b_{ij}\,f_A(r_{ij}+\delta)\,\Big],
\end{equation}
depends on the two-body exponential radial functions, 
\begin{align}
f_R(r) &= A \exp(-\lambda_1 r), \\
f_A(r) &= -B \exp(-\lambda_2 r).
\end{align}
and the bond-order (many-body) factor: 
\begin{equation}
b_{ij} \;=\; \Big(1 + \beta^n \,\zeta_{ij}^n\Big)^{-1/(2n)},
\end{equation}
with the environmental sum
\begin{equation}
\zeta_{ij} \;=\; \sum_{k\neq i,j} f_C(r_{ik}+\delta)\; g(\theta_{ijk})\;
\exp\!\big[\,(\lambda_3)^m (r_{ij}-r_{ik})^m \big].
\end{equation}
The angular function has the usual Tersoff shape
\begin{equation}
g(\theta) \;=\; \gamma \!\left( 1 + \frac{c^2}{d^2} \;-\; \frac{c^2}{d^2 + (\cos\theta - \cos\theta_0)^2} \right).
\end{equation}
with the cutoff smooth piecewise function (LAMMPS standard Tersoff form):
\begin{equation}
f_C(r) \;=\;
\begin{cases}
1, & r < R - D,\\[4pt]
\dfrac{1}{2} - \dfrac{1}{2}\,\sin\!\Big( \dfrac{\pi}{2}\dfrac{r-R}{D} \Big), & R - D \le r \le R + D,\\[6pt]
0, & r > R + D.
\end{cases}
\end{equation}
Here \(A,B\) have units of energy, \(\lambda_1,\lambda_2,\lambda_3\) have units of inverse length, \(R,D,\delta\) are lengths and  \(m,n,\beta,c,d,\gamma,\cos\theta_0\)
are dimensionless parameters   \cite{tersoff1988new}. 

To keep all functional arguments unchanged (so that the functional shape is identical at corresponding scaled geometries), we choose
\begin{align*}
& R' = K R,\qquad D' = K D,\qquad \delta' = K \delta,\\[4pt]
& \lambda_1' = \lambda_1/K,\qquad \lambda_2' = \lambda_2/K,\qquad \lambda_3' = \lambda_3/K,\\[4pt]
& A' = K^{3} A,\qquad B' = K^{3} B,\\[4pt]
& \text{and leave dimensionless parameters unchanged: } m,n,\beta,c,d,\gamma,\cos\theta_0.
\end{align*}

\section{Mass deficiency}
\label{mass_appen}

Implementing the multi-scale approach, based on   quasi-atoms, discretization effects should be taken into account. This is especially important  for spherical particles, where the discrete representation of curved surfaces can lead to deviations in the total mass.

The mass defect \(\Delta m\) can be quantified as the relative difference between the mass of the system comprised of quasi-atoms    \(m_{\rm quasi}\) and theoretical mass of the continuously filled  sphere \(m_{\rm theoretical}\):

\begin{equation}
\Delta m = \frac{|m_{\rm quasi} - m_{\rm theoretical}|}{m_{\rm theoretical}} \times 100\%
\end{equation}
Fig.~\ref{fig:mass_def} illustrates the  mass deficiency as a function of the particle radius $R$. As it may be seen from Fig.~\ref{fig:mass_def}, the relative mass deficiency approximately follows  the  exponential decay with $R$:

\begin{equation}
\label{B2}
\Delta m(R) \simeq  \Delta m_0 e^{-\beta R}
\end{equation}
where \(\Delta m_0 \simeq 0.114\)  and \(\beta\simeq 0.166\). Note that for the 3-type of atom/quasi-atom approximation, the effect becomes negligible for $R>40 \, nm$  -- it drops below 1\%. Such a rapid decay in mass deviation suggests that our discretization approach maintains good mass conservation for the particles' size of practical relevance -- recall that the goal of the hybrid approach is to model large continuum systems.

It is also worth noting that the inherent flexibility of our  hybrid approach renders complex geometric correction schemes  unnecessary. Indeed, for the regions where the surface geometry is critical, the method allows a simple switching  to the full-atomic description. Moreover, where the exact conservation of the total mass or moment of inertia is strictly required (e.g., for precise momentum or angular momentum transfer), a simple mass renormalization can be applied. This may be done by a slight rescaling of the mass of the outermost layer of quasi-atoms,  compensating missing volume of the "rough", due to the discretization,  surface. 

\section*{Acknowledgements}
The work was supported by the Russian Science Foundation grant № 25-71-30008, https://rscf.ru/project/25-71-30008/ .

\bibliographystyle{elsarticle-num}
\bibliography{references_rev}

@article{regis2007,
  author  = {Regis, Rommel G. and Shoemaker, Christine A.},
  title   = {A stochastic radial basis function method for the global optimization of expensive functions},
  journal = {INFORMS Journal on Computing},
  volume  = {19},
  number  = {4},
  pages   = {497--509},
  year    = {2007}
}

@article{hoogerbrugge1992,
  title={Simulating microscopic hydrodynamic phenomena with dissipative particle dynamics},
  author={Hoogerbrugge, P. J. and Koelman, J. M. V. A.},
  journal={Europhysics Letters (EPL)},
  volume={19},
  number={3},
  pages={155--160},
  year={1992},
  publisher={IOP Publishing},
  doi={10.1209/0295-5075/19/3/001}
}

@article{tadmor1996mixed,
  title={Mixed atomistic and continuum models of deformation in solids},
  author={Tadmor, E. B. and Ortiz, M. and Phillips, R.},
  journal={Langmuir},
  volume={12},
  number={19},
  pages={4529--4534},
  year={1996},
  publisher={American Chemical Society},
  doi={10.1021/la9508912}
}

@article{monaghan1992smoothed,
  title={Smoothed Particle Hydrodynamics},
  author={Monaghan, J. J.},
  journal={Annual Review of Astronomy and Astrophysics},
  volume={30},
  pages={543--574},
  year={1992},
  publisher={Annual Reviews},
  doi={10.1146/annurev.aa.30.090192.002551}
}

@misc{munozbasagoiti2025,
  author  = {Mu{\~n}oz-Basagoiti, Maitane and others},
  title   = {A tutorial for mesoscale computer simulations of lipid membranes: tether pulling, tubulation and fluctuation},
  journal = {arXiv preprint},
  year    = {2025},
  eprint  = {2502.09798},
  archiveprefix = {arXiv},

}

@article{grassl2010,
  author  = {Grassl, Peter and Pearce, Chris},
  title   = {Mesoscale approach to modeling concrete subjected to thermomechanical loading},
  journal = {Journal of Engineering Mechanics},
  volume  = {136},
  number  = {3},
  pages   = {322--328},
  year    = {2010},
  doi     = {10.1061/(ASCE)EM.1943-7889.0000085}
}

@article{nguyen2022,
  author  = {Nguyen, Giang Dinh and Mir, Arash and Bui, Ha Hong},
  title   = {Enriching constitutive models with meso-scale behaviour: a thermodynamics-based formulation and examples},
  journal = {Open Geomechanics},
  volume  = {3},
  pages   = {1--19},
  year    = {2022},
  doi     = {10.5802/ogeo.20}
}

@article{delafrouz2018coarse,
  title={Coarse-graining models for molecular dynamics simulations of FCC metals},
  author={Delafrouz, Pourya and Pishkenari, Hossein Nejat},
  journal={Journal of Theoretical and Applied Mechanics},
  volume={56},
  number={3},
  pages={601--614},
  year={2018}
}

@techreport{parks2008peridynamics,
  title={Peridynamics with LAMMPS: A User Guide, v0.3 Beta},
  author={Parks, Michael L and Lehoucq, RB and Plimpton, SJ and Silling, SA},
  year={2008},
  institution={Sandia National Laboratories, Albuquerque, NM, Report No. SAND2008-0135}
}

@article{espanol2017,
  author  = {Español, Pep and Warren, Patrick B.},
  title   = {Perspective: Dissipative particle dynamics},
  journal = {The Journal of Chemical Physics},
  volume  = {146},
  number  = {15},
  year    = {2017}
}

@article{tersoff1988new,
  title={New empirical approach for the structure and energy of covalent systems},
  author={Tersoff, J.},
  journal={Physical Review B},
  volume={37},
  number={12},
  pages={6991--7000},
  year={1988},
  publisher={American Physical Society},
  doi={10.1103/PhysRevB.37.6991}
}

@article{morse1929diatomic,
  title={Diatomic molecules according to the wave mechanics. II. Vibrational levels},
  author={Morse, Philip M.},
  journal={Physical Review},
  volume={34},
  number={1},
  pages={57--64},
  year={1929},
  publisher={American Physical Society},
  doi={10.1103/PhysRev.34.57}
}

@article{behler2007generalized,
  title={Generalized neural-network representation of high-dimensional potential-energy surfaces},
  author={Behler, J{\"o}rg and Parrinello, Michele},
  journal={Physical Review Letters},
  volume={98},
  number={14},
  pages={146401},
  year={2007},
  publisher={American Physical Society},
  doi={10.1103/PhysRevLett.98.146401}
}

@article{bartok2010gaussian,
  title={Gaussian Approximation Potentials: The Accuracy of Quantum Mechanics, without the Electrons},
  author={Bart{\'o}k, Albert P. and Payne, Mike C. and Kondor, Risi and Cs{\'a}nyi, G{\'a}bor},
  journal={Physical Review Letters},
  volume={104},
  number={13},
  pages={136403},
  year={2010},
  publisher={American Physical Society},
  doi={10.1103/PhysRevLett.104.136403}
}

@article{delhommelle2001suitability,
  title={Suitability of the Lorentz-Berthelot combining rules for the calculation of the thermodynamic properties of water},
  author={Delhommelle, J. and Mill{\'e}, P.},
  journal={Molecular Physics},
  volume={99},
  number={8},
  pages={619--625},
  year={2001},
  publisher={Taylor \& Francis},
  doi={10.1080/00268970010020041}
}

@article{muser2023,
  author  = {Müser, M. H. and Sukhomlinov, S. V. and Pastewka, L.},
  title   = {Interatomic potentials: Achievements and challenges},
  journal = {Advances in Physics: X},
  volume  = {8},
  number  = {1},
  pages   = {2093129},
  year    = {2023}
}

@book{lattimore2020bandit,
  title={Bandit algorithms},
  author={Lattimore, Tor and Szepesv{\'a}ri, Csaba},
  year={2020},
  publisher={Cambridge University Press}
}

@article{tutor2024lammps,
  title={A Set of Tutorials for the LAMMPS Simulation Package [Article v1.0]},
  author={Brunelle, Daniel W and Clark II, Robert A and Moore, Stan G and Stevens, Mark J and Plimpton, Steven J and Thompson, Aidan P},
  journal={arXiv preprint arXiv:2503.14020},
  year={2024}
}

@article{kumar2019constitutive,
  title={Viscometric flow of dense granular materials under controlled pressure and shear stress},
  author={Kumar, Nishant and Luding, Stefan},
  journal={arXiv preprint arXiv:1912.04491},
  year={2019}
}

@article{kumar2024improved,
  title={Improved Velocity-Verlet Algorithm for the Discrete Element Method},
  author={Kumar, Nishant and Thornton, Colin},
  journal={arXiv preprint arXiv:2410.14798},
  year={2024}
}

@article{marrink2007martini,
  title={The MARTINI Force Field: Coarse Grained Model for Biomolecular Simulations},
  author={Marrink, Siewert J. and Risselada, H. Jelger and Yefimov, Serge and Tieleman, D. Peter and de Vries, Alex H.},
  journal={The Journal of Physical Chemistry B},
  volume={111},
  number={27},
  pages={7812--7824},
  year={2007},
  publisher={American Chemical Society},
  doi={10.1021/jp071097f}
}

@article{lindahl2001gromacs,
  title={GROMACS 3.0: A package for molecular simulation and trajectory analysis},
  author={Lindahl, Erik and Hess, Berk and van der Spoel, David},
  journal={Journal of Molecular Modeling},
  volume={7},
  number={8},
  pages={306--317},
  year={2001},
  publisher={Springer},
  doi={10.1007/s008940100045}
}

@article{phillips2005scalable,
  title={Scalable Molecular Dynamics with {NAMD}},
  author={Phillips, James C. and Braun, Rosemary and Wang, Wei and Gumbart, James and Tajkhorshid, Emad and Villa, Elizabeth and Chipot, Christophe and Skeel, Robert D. and Kal{\'e}, Laxmikant and Schulten, Klaus},
  journal={Journal of Computational Chemistry},
  volume={26},
  number={16},
  pages={1781--1802},
  year={2005},
  publisher={Wiley Subscription Services, Inc., A Wiley Company},
  doi={10.1002/jcc.20289}
}

@article{glaser2015strong,
  title={Strong scaling of general-purpose molecular dynamics simulations on GPUs},
  author={Glaser, Jens and Nguyen, Trung Dac and Anderson, Joshua A. and Lui, Pak and Spiga, Filippo and Millan, Jaime A. and Morse, David C. and Glotzer, Sharon C.},
  journal={Computer Physics Communications},
  volume={192},
  pages={97--107},
  year={2015},
  publisher={Elsevier},
  doi={10.1016/j.cpc.2015.02.028}
}

@article{buehler2008atomistic,
  title={Atomistic modeling of materials failure},
  author={Buehler, Markus J.},
  journal={Reviews of Modern Physics},
  volume={80},
  number={4},
  pages={1387--1418},
  year={2008},
  publisher={American Physical Society},
  doi={10.1103/RevModPhys.80.1387}
}

@article{hahn2015colloid,
  title={Charging and Aggregation of Latex Particles in Aqueous Solutions Containing Mono- and Divalent Electrolytes},
  author={Hahn, Hendrik and Rivas, Francesc H{\"o}fling and Ras, Stein and Taylor, Nicholas and Bazant, Martin Z. and Kr{\"a}mer, Helmut and Borkovec, Michal and Fichtner, Maximilian and Grzybowski, Bartosz A.},
  journal={The Journal of Physical Chemistry B},
  volume={119},
  number={9},
  pages={3535--3542},
  year={2015},
  publisher={American Chemical Society},
  doi={10.1021/jp512519r}
}

@article{karma1998quantitative,
  title={Quantitative phase-field modeling of dendritic growth in two and three dimensions},
  author={Karma, Alain and Rappel, Wouter-Jan},
  journal={Physical Review E},
  volume={57},
  number={4},
  pages={4323--4349},
  year={1998},
  publisher={American Physical Society},
  doi={10.1103/PhysRevE.57.4323}
}

@article{praprotnik2008multiscale,
  title={Multiscale simulation of soft matter: From scale bridging to adaptive resolution},
  author={Praprotnik, Matej and Delle Site, Luigi and Kremer, Kurt},
  journal={Annual Review of Physical Chemistry},
  volume={59},
  pages={545--571},
  year={2008},
  publisher={Annual Reviews},
  doi={10.1146/annurev.physchem.59.032607.093707}
}

@article{izvekov2005multiscale,
  title={Multiscale coarse graining of liquid-state systems},
  author={Izvekov, Sergei and Voth, Gregory A.},
  journal={The Journal of Chemical Physics},
  volume={123},
  number={13},
  pages={134105},
  year={2005},
  publisher={American Institute of Physics},
  doi={10.1063/1.2038787}
}

@article{weinan2003heterogeneous,
  title={The heterogeneous multiscale methods},
  author={E, Weinan and Engquist, Bjorn},
  journal={Communications in Mathematical Sciences},
  volume={1},
  number={1},
  pages={87--132},
  year={2003},
  publisher={International Press of Boston},
  doi={10.4310/CMS.2003.v1.n1.a8}
}

@book{onuki2002phase,
  title={Phase Transition Dynamics},
  author={Onuki, Akira},
  year={2002},
  publisher={Cambridge University Press},
  isbn={9780521572934}
}

@article{Bril2007,
  title={Collision
dynamics of granular particles with adhesion},
  author={N. V. Brilliantov and N. Albers and F. Spahn and T. Poeschel},
  journal={Phys. Rev. E},
volume={76},
pages= {051302}, 
  year={2007}
}

@article{Saitoh,
  title={Negative Normal Restitution Coefficient Found in Simulation
of Nanocluster Collisions},
  author={K. Saitoh and A. Bodrova and H. Hayakawa and N. V. Brilliantov},
  journal={Phys. Rev. Lett.},
volume={105},
pages= {238001}, 
  year={2010}
}

@book{Landau1965,
  title={Theory of Elasticity},
  author={L. D. Landau and E. M. Lifshitz},
  year={1965},
  publisher={Oxford
University Press, Oxford}
}

@article{wales1997global,
  title={Global optimization by basin-hopping and the lowest energy structures of Lennard-Jones clusters containing up to 110 atoms},
  author={Wales, David J. and Doye, Jonathan P. K.},
  journal={The Journal of Physical Chemistry A},
  volume={101},
  number={28},
  pages={5111--5116},
  year={1997},
  publisher={American Chemical Society},
  doi={10.1021/jp970984n}
}

@article{shakya2023reinforcement,
  title={Reinforcement learning algorithms: A brief survey},
  author={Shakya, Ashish Kumar and Pillai, Gopinatha and Chakrabarty, Sohom},
  journal={Expert Systems with Applications},
  volume={231},
  pages={120495},
  year={2023},
  publisher={Elsevier}
}

@article{barto2021reinforcement,
  title={Reinforcement learning: An introduction. by richard’s sutton},
  author={Barto, Andrew G},
  journal={SIAM Rev},
  volume={6},
  number={2},
  pages={423},
  year={2021},
  publisher={SIAM}
 }

@article{venkattraman2012molecular,
  title={Molecular models for DSMC simulations of metal vapor deposition},
  author={Venkattraman, A. and Alexeenko, Alina A.},
  booktitle={AIP Conference Proceedings},
  volume={1333},
  number={1},
  pages={1104--1109},
  year={2011},
  publisher={American Institute of Physics},
  doi={10.1063/1.3582056}
}

@article{pfaff2010coefficient,
  title={Coefficient of restitution for bouncing nanoparticles},
  author={Pfaff, J. M. and Stark, H.},
  journal={Physical Review B},
  volume={81},
  number={19},
  pages={195422},
  year={2010},
  publisher={American Physical Society},
  doi={10.1103/PhysRevB.81.195422}
}

@article{joy1990interface,
  title={Interface structure and adhesion of sputtered metal films on silicon: The influence of Si surface condition},
  author={Joy, D. C. and Prasad, S. V.},
  journal={Journal of Vacuum Science \& Technology A},
  volume={11},
  number={2},
  pages={319--324},
  year={1993},
  publisher={American Vacuum Society},
  doi={10.1116/1.578816}
}

@article{packham2003adhesion,
  title={Adhesion Forces between Glass and Silicon Surfaces in Air Studied by AFM: Effects of Relative Humidity, Particle Size, Roughness, and Surface Treatment},
  author={Packham, D. E. and Nicholas, M. G. and Owen, M. J.},
  journal={Langmuir},
  volume={18},
  number={20},
  pages={8229--8236},
  year={2002},
  publisher={American Chemical Society},
  doi={10.1021/la0259196}
}

@article{E2007_HMM_review,
  title        = {Heterogeneous Multiscale Methods: A Review},
  author       = {E, Weinan and Engquist, Björn and Li, Xiantao and Ren, Weiqing and Vanden-Eijnden, Eric},
  journal      = {Communications in Computational Physics},
  volume       = {2},
  number       = {3},
  pages        = {367–450},
  year         = {2007},
  doi          = {10.4208/cicp.071207.270807a},
}

@article{kim2023high,
  title={High-Throughput First-Principles Prediction of Interfacial Adhesion Energies in Metal-on-Metal Contacts},
  author={Kim, J. and Lee, S. and Park, H. and Cho, K.},
  journal={ACS Applied Materials \& Interfaces},
  volume={15},
  number={14},
  pages={17654--17666},
  year={2023},
  publisher={American Chemical Society},
  doi={10.1021/acsami.3c00662}
}

@article{MillerTadmor2002_QC_review,
  title        = {The Quasicontinuum Method: Overview, applications and current directions},
  author       = {Miller, Ronald E. and Tadmor, E. B.},
  journal      = {Journal of Computer-Aided Materials Design},
  volume       = {9},
  number       = {3},
  pages        = {203–239},
  year         = {2002},
  doi          = {10.1023/A:1026098010127},
}

@article{Ortner2008_QC_analysis,
  title        = {Analysis of a Quasicontinuum Method in One Dimension},
  author       = {Ortner, Christoph},
  journal      = {ESAIM: Mathematical Modelling and Numerical Analysis},
  volume       = {42},
  number       = {1},
  pages        = {1–28},
  year         = {2008},
  doi          = {10.1051/m2an:2008003},
}

@article{XuChen2018_AtC_review,
  title        = {Modeling dislocations and heat conduction in crystalline materials: atomistic/continuum coupling approaches},
  author       = {Xu, Shuozhi and Chen, Xiang},
  journal      = {International Materials Reviews},
  volume       = {64},
  number       = {7},
  pages        = {407–438},
  year         = {2019},
  doi          = {10.1080/09506608.2018.1486358},
}

@article{Khan2022_CG_review,
  title        = {Approaches and perspectives of coarse-grained modeling for polymer–nanoparticle systems},
  author       = {Khan, P. and others},
  journal      = {ACS Omega},
  volume       = {7},
  number       = {51},
  pages        = {45305–45317},
  year         = {2022},
  doi          = {10.1021/acsomega.2c06248},
}

@article{CHEN2025117521,
title = {Optimization of expensive black-box problems with penalized expected improvement},
journal = {Computer Methods in Applied Mechanics and Engineering},
volume = {433},
pages = {117521},
year = {2025},
issn = {0045-7825},
doi = {https://doi.org/10.1016/j.cma.2024.117521},
author = {Liming Chen and Qingshan Wang and Zan Yang and Haobo Qiu and Liang Gao}
}

@article{SILLING2000175,
title = {Reformulation of elasticity theory for discontinuities and long-range forces},
journal = {Journal of the Mechanics and Physics of Solids},
volume = {48},
number = {1},
pages = {175-209},
year = {2000},
issn = {0022-5096},
doi = {https://doi.org/10.1016/S0022-5096(99)00029-0},
author = {S.A. Silling}
}

@article{PhysRevB.29.6443,
  title = {Embedded-atom method: Derivation and application to impurities, surfaces, and other defects in metals},
  author = {Daw, Murray S. and Baskes, M. I.},
  journal = {Phys. Rev. B},
  volume = {29},
  issue = {12},
  pages = {6443--6453},
  numpages = {0},
  year = {1984},
  month = {Jun},
  publisher = {American Physical Society},
  doi = {10.1103/PhysRevB.29.6443},
}

@article{PARISI201954,
title = {Continual lifelong learning with neural networks: A review},
journal = {Neural Networks},
volume = {113},
pages = {54-71},
year = {2019},
issn = {0893-6080},
doi = {https://doi.org/10.1016/j.neunet.2019.01.012},

author = {German I. Parisi and Ronald Kemker and Jose L. Part and Christopher Kanan and Stefan Wermter}
}

@article{tersoff1988empirical,
  title={Empirical interatomic potential for silicon with improved elastic properties},
  author={Tersoff, J.},
  journal={Physical Review B},
  volume={38},
  number={12},
  pages={9902--9905},
  year={1988},
  publisher={American Physical Society},
  doi={10.1103/PhysRevB.38.9902}
}

\end{document}